\documentclass[12pt]{article}
\usepackage{macros2elang}
\usepackage{amssymb,epsf,graphicx
}
\newcommand{\E}{\epsilon}
\newcommand{\be}{\begin{equation}}
\newcommand{\ee}{\end{equation}}
\newcommand{\bea}{\begin{eqnarray}}
\newcommand{\eea}{\end{eqnarray}}
\newcommand{\rmd}{\mathrm{d}}
\newcommand{\rme}{\mathrm{e}}
\newcommand{\nn}{\nonumber}

\newcommand{\eq}[1]{(\ref{#1})}
\newcommand{\Eq}[1]{Eq.\hspace{0.55ex}(\ref{#1})}
\newcommand{\Eqs}[1]{Eqs.\hspace{0.55ex}(\ref{#1})}
\newcommand{\Fig}[1]{Fig.\hspace{0.55ex}\ref{#1}}
\newcommand{\half}{\frac12}
\newcommand{\Sec}[1]{Sect.\hspace{0.55ex}\ref{#1}}

\begin{document}

\begin{titlepage}
\noindent
\renewcommand{\thefootnote}{\fnsymbol{footnote}}
\parbox{4.89cm}{\epsfxsize=4.89cm
\epsfbox{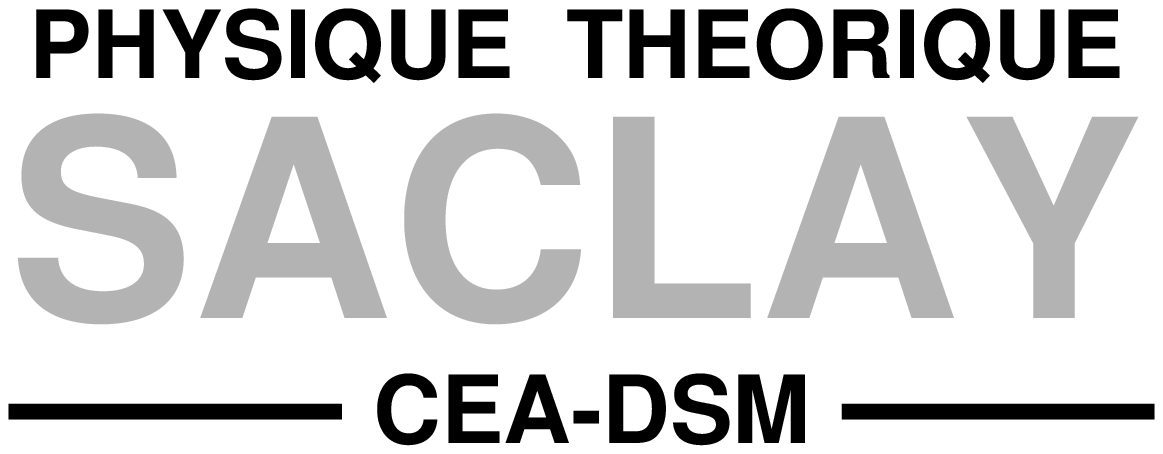}}%
\hspace{0.5cm}%
\parbox{1.85cm}{\epsfysize=1.85cm \epsfbox{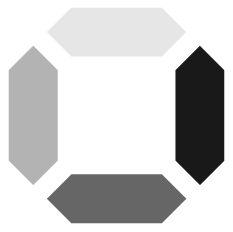}}%
\hfill
\begin{minipage}{1cm}
\rightline{cond-mat/9807160}
\rightline{Saclay Preprint T/98065}
\rightline{Uni Essen preprint}
\end{minipage}
\vfill
\vskip5mm
\centerline{\sffamily\bfseries\Large 
Large Orders for Self-Avoiding Membranes}
\vfill
\centerline{\bf\large Fran\c cois David%
\footnote{Email: david@spht.saclay.cea.fr}%
\footnote{Physique Th\'eorique CNRS}
}
\smallskip
\centerline{\small SPhT, C.E.A. Saclay, 91191 Gif-sur-Yvette Cedex, France}
\vskip1cm
\centerline{\bf\large Kay J\"org Wiese%
\footnote{Email: wiese@theo-phys.uni-essen.de}}
\smallskip
\centerline{\small Fachbereich Physik, Universit\"at GH Essen,  45117 Essen,
Germany}

\vspace*{-5mm}
\vfill

\begin{abstract}
\noindent
We derive the large order behavior of the perturbative expansion for the
continuous model of tethered self-avoiding membranes.
It is controlled by a classical configuration for an
effective potential in bulk space, which is the analog of the Lipatov instanton, solution of a highly non-local equation.
The $n$-th order is shown to have factorial growth as
$(-\mathrm{cst})^n\,(n!)^{(1-\epsilon/D)}$, where $D$ is the ``internal"
dimension of the membrane and $\epsilon$ the engineering dimension of the coupling constant for self-avoidance.
The instanton is calculated within a variational approximation, which is 
shown to become exact in the limit of large dimension $d$  of bulk space.
This  is the starting point of a systematic $1/d$ expansion.
As a consequence, the $\epsilon$-expansion of 
self-avoiding membranes has a factorial growth,
like the $\epsilon$-expansion of polymers and standard critical phenomena,
suggesting Borel summability.
Consequences for the applicability of the 2-loop calculations are examined.
\end{abstract}

\vspace{-5mm}
\vfill

\centerline{Submitted to Nucl.~Phys.~{\bf B}}

\vfill

\end{titlepage}\addtocounter{page}{1}
\renewcommand{\thefootnote}{\arabic{footnote}}
\setcounter{footnote}{0}

\section{Introduction}\label{s.intro}

Flexible polymerized tethered membranes (also called polymerized
membranes
or tethered membranes) exhibit fascinating statistical properties
\cite{NelPel87,Jerus89}. 
Tethered membranes with only short-range  repulsive
interactions (``self-avoidance'') may be seen as
 the 2-dimensional analog of  polymer chains in
a good solvent. They are expected to exist in two phases:
(1) a flat (or rigid) phase with an infinite persistence length but
non-classical elastic properties, 
not found for polymers which have a finite persistence length
\cite{PelLei85,PacKarNel88,DavGui88},
and 
(2) a crumpled phase similar to that of polymers
\cite{KanKarNel86,KanKarNel87}.
At variance with polymers, the crumpled phase is always
 swollen  (whatever the dimension $d$ of space is) and has a configuration exponent
$\nu$ larger than the mean-field exponent
$\nu_{\scriptscriptstyle\mathrm{mf}}=0$, whereas 
for polymers it is well known that for $d\ge 4$ one has
$\nu=\nu_{\scriptscriptstyle\mathrm{mf}}=1/2$ while for $d<4$ 
one has $\nu>\nu_{\scriptscriptstyle\mathrm{mf}}$.
This follows from a simple dimensional argument, which shows that self-avoidance
is always relevant (in the renormalization group sense)
\cite{KanKarNel86}. 

The properties of the crumpled swollen phase have been much studied theoretically.
In particular, a systematic framework for renormalization group  (RG)
calculations has been proposed in 
\cite{AroLub88,KarNel87,KarNel88}.
It consists in:
(1) a self-avoiding membrane model (SAM) which is a non-trivial extension of
the continuous Edwards model for polymers
\cite{Edw65,CloJan90},
and 
(2) a renormalization procedure which generalizes the direct renormalization
\cite{Clo81} for the Edwards model. 
For 2-dimensional
 tethered membranes, the upper critical dimension $d_{\mathrm{uc}}$
is infinite, but the SAM model can be  extended to membranes with non-integer internal dimension $D<2$, which have a finite $d_{\mathrm{uc}}$.
Then the framework of 
\cite{AroLub88,KarNel87} 
allows to perform a double expansion both in $D$ (the internal
dimension of the membrane) and $d$ (the dimension of bulk space)
to study the physical case ($D=2$, $d<\infty$) of a 2-dimensional membrane
in finite dimension $d$.
The starting point for the $\epsilon$-expansion (characterized by the
dimension $D_0<2$ for the membrane) can be chosen arbitrarily.
This provides an additional parameter for the expansion, which can be used to
optimize the calculations
\cite{Hwa90,WieDav97}.

This approach, which amounts to perform perturbative renormalization group
calculations for $D$-dimensional membranes, has raised considerable challenges.
Whereas for polymers (i.e.\ when D=1) the Edwards model can be mapped onto the 
local $\Phi^4$ field theory with $n=0$ components
\cite{Gen72}, and direct renormalization
is equivalent to the standard renormalization (in the MS scheme)
\cite{BenMah86},
this equivalence does not hold for $D\ne 1$.
It was finally shown in 
\cite{DavDupGui96,DavDupGui97}
that the perturbative RG calculations are mathematically consistent, by proving
that the SAM model is renormalizable to all orders in perturbation theory.
If perturbative calculations are simple at first order (with 
some subtle points
\cite{Dup87}), they present considerable difficulties at second order,
and require a lot of analytical and numerical work
\cite{DavWie96,WieDav97}.

An important issue is to understand if these calculations make sense
beyond perturbation theory, or if non-perturbative effects destroy
the consistency of the approach. 
A first step is to understand the large order behavior of perturbation theory
for the SAM model. 
To our knowledge, nothing was known about this problem 
up to now, except  for $D=1$, where one can use the equivalence to the 
$n\to0$ limit of $\Phi^4$-theory, for which a non-trivial solution of
the equation of motion at negative coupling constant, the so-called instanton,
governs the large order behavior \cite{Lip76,Zin82};
however, this analogy does not provide a physical picture of which
polymer-configurations dominate the large orders.
In this paper we shall formulate the problem of the large order behavior
for the Edwards model, in a way which is directly applicable both to
polymers and to membranes.  
\medskip

Let us summarize the organization and the results of this paper.
In \Sec{s.2} we recall the definition of the SAM model.
Then we develop a general semiclassical argument to compute the large
orders of the perturbative series for the SAM model.
Using the formulation of the SAM model as a model of a ``phantom" membrane
(without self-avoidance) in a random imaginary external potential
$V$, we show that the large orders are controlled by a real classical configuration for this potential $V$, which is the analog for SAM of the
instanton for $\Phi^4$.
This ``SAM instanton" potential $V$ is the extremum of a non-local
functional $\mathcal{S}[V]$, which cannot be calculated exactly.
We obtain the general form for the asymptotics of the term of order $n$,
which is
\begin{equation}
n^{d/2}\,\left(-\,\mathcal{C}\right)^n\,(n!)^{1-\epsilon/D} \ ,
\label{1steq}
\end{equation}
where $D$ is the internal dimension of the membrane, $d$  the dimension of 
bulk space, $\epsilon=2D-d(2-D)/2$  the engineering dimension for the 
self-avoidance coupling in the SAM model and $\mathcal{C}$ a positive
constant depending on $D$ and $\epsilon$ (or $d$).
This behavior is universal: the constant $\mathcal{C}$ 
obtained from the instanton does not depend on the
internal shape or topology of the membrane.

In \Sec{s.3} we show that for polymers ($D=1$) our results fully agree with 
the classical results on the large orders for the $\Phi^4$ theory, as obtained
from instanton calculus.
We then give a simple physical interpretation of our instanton for the SAM 
model: the SAM instanton describes the metastable equilibrium configuration
of a membrane submitted to two competing forces, a contact attractive
interaction and a global entropic repulsion due to thermal fluctuations.

The SAM instanton equations are in general not solvable.
In \Sec{s.4} we propose a Gaussian variational approximation scheme
and find the instanton potential $V$ and the large order constant $\mathcal{C}$
within this approximation. By construction, this is a lower bound on the 
exact result.

The variational results are discussed in \Sec{s.5}.
They qualitatively agree with the exact results for polymers ($D=1$).
We estimate the ``optimal order" for the $\epsilon$-expansion,
and we argue that the estimates for the configuration exponent $\nu$ obtained
from second order calculations \cite{DavWie96,WieDav97}
are reasonable, especially for large bulk dimension $d$.
Finally we observe that they imply that the perturbative series for the SAM
model should become ``quasi-convergent" in the limit $d\to\infty$.

In \Sec{s.6} we study more carefully the validity of the variational
approach.
We show that it becomes exact in the limit of large space  dimension
$d\to\infty$. Technically one has to keep $\epsilon$ fixed, that is to let
the internal dimension $D$ of the membrane go to $2$.
To go beyond the leading order, we construct a systematic perturbative expansion around the
variational solution, and show that this expansion can be organized
as a systematic expansion in powers of $1/d$, by resumming infinite classes
of diagrams.
We compute explicitly the first correction in $1/d$ for the instanton $V$ and
the large order constant $\mathcal{C}$.
In particular, we find that along the critical line $\epsilon=0$, the
instanton potential $V$ has $\log(\epsilon)$ singularities, but that
the large order constant $\mathcal{C}$ is finite.

In \Sec{s.7} we present our conclusions and discuss open problems.

\section{Large orders and instantons for the SAM model}
\label{s.2}
\subsection{The SAM model}

We consider a  $D$-dimensional
manifold ${\cal M}$ with size $L$ and volume ${\cal V}=L^{D}$
(typically the
$D$-dimensional torus $\mathrm{T}_D=[0,L]^{D}$) in $d$-dimensional Euclidean 
bulk space.
A configuration of the manifold is described by the continuous
function
\be
x\in {\cal M}\ \rightarrow\ \vec r(x) \in \mathbb{R}^{d} \ .
    \label{mapping}
\ee
The partition function is
\be
{\cal Z}[b;L]\ =\ \int {\cal D}[\vec r]\ e^{-{\cal H}[\vec r;b,L]}
	\label{partition}
\ee
with the Hamiltonian
\be
{\cal H}[\vec r;b,L]\ =\ \int_{L}\,\rmd^{D}\!x\,{1\over2}(\nabla\vec
r(x))^{2}+{b\over 2}\,\int_{L}\rmd^{D}\!x\,\int_{L}\rmd^{D}\!y\,\delta^{d}(\vec 
r(x)-\vec
r(y))	
	\label{hamiltonian} \ .
\ee
$b>0$ is the repulsive 2-point interaction coupling which describes
self-avoidance.
The functional integration measure ${\cal D}[\vec
r]=\prod_{x}\rmd^{d}\vec
r(x)/{\cal Z}_{0}$ is normalized such
that the partition function of the free Gaussian
manifold ${\cal Z}[b=0,L]=1$.
The canonical dimensions of $x$, $\vec r$ and $b$ are
\be
	[x]\ =\ -1\quad,\quad[\vec r]\ =\ {D-2\over2}\quad,\quad[b]\ =\
	2D-{d\,(2-D)\over 2}\ =\ \E
	\label{canonical}
\ .
\ee

We know that the partition function ${\cal Z}$ and
expectation values
of in bulk space
translationally invariant  operators
$\langle \cal O\rangle$ are well defined as a perturbative
series in the coupling constant $b$, as long as the dimension of the
manifold $D$ is less than 2 and as long as $\E$ is positive
\be
	0\ <\ D\ < 2 \ ,\qquad\E\ >\ 0
	\label{Dle2}
\ee
and provided that the size $\cal V$ of the manifold  is finite%
\footnote{Strictly  speaking, for $0<\E<D$, the first terms in
the expansion of the
partition function suffer from short-distance divergences.
These   Ultra-Violet (UV) singularities
are  finite in number
and can be recast
in a counterterm proportional to the volume $\cal V$ of the
manifold.
Moreover they disappear in physical observables; they are thus of
 no physical significance and unimportant for the
purpose of this paper.}.
For $\E>0$, perturbation theory suffers
from strong Infra-Red (IR) divergences when the size of the manifold
becomes large $L\to\infty$, which signal a breakdown of mean-field
theory and the appearance of anomalous dimensions in the  scaling
properties
of large self-avoiding manifolds.

For $\E\to 0$, physical UV singularities appear at all orders
in perturbation expansion. They can be absorbed in a
renormalization of the coupling constant $b$ and a rescaling of bulk
space (wave function renormalization)
\be
	b\ = \ b_{R}\,Z_{b}(b_{R})\ ,\qquad\vec r\ = \
	Z^{1/2}(b_{R})\,\vec r_{R}
	\label{renorm}
\ee
so that physical observables are UV finite at $\E=0$ when
expressed in terms of renormalized quantities $b_{R}$ and $\vec
r_{R}$.
The theory, although non-local in the internal $x$-space of the
membrane, is renormalizable 
\cite{DavDupGui96,DavDupGui97}.
As a consequence, it is possible to write renormalization group
equations which encode how the effective renormalized coupling flows
with the
length scale, and in particular with the size $L$ of the membrane.
The large $L$ behavior of the membrane is governed by an IR stable
fixed point $b_{R}^{*}$, which can in principle be calculated order
by order in perturbation theory as a series in $\E$, similar to
the celebrated Wilson-Fisher $\E$-expansion.
In practice, such calculations are very difficult, and have been
performed only up to second order 
\cite{DavWie96,WieDav97}.

\subsection{Instanton calculus}
By dimensional analysis, the partition function (\ref{partition})
only depends  on the dimensionless coupling constant
\be
g\ =\ b\,L^{\E}
\label{dimlessg}
\ee
via
\be
{\cal Z}[b;L]\ =\ {\cal Z}[g,L=1]\ =\ {\cal Z}[g]
\label{dimlessZ}
\ee
and is defined as a series
\be
{\cal Z}[g]\ =\ \sum_{n=0}^{\infty}\,z_{n}\,g^{n}
\label{defzn}
\ .
\ee
Of course, ${\cal Z}[g]$ also depends  on the shape of the manifold
$\cal M$.

Let us assume that ${\cal Z}[g]$ is analytic around the origin for
$-\pi<\arg(g)<\pi$, and has a discontinuity along the negative real
axis. This assumption is natural, since for  $g<0$, the membrane
is in a collapsed state and the perturbation expansion is performed
around an unstable classical state.
Then we can write $z_{n}$ as a dispersion integral
\be
z_n\ =\ \oint\,{\rmd g\over 2i\pi}\, g^{-n-1}\,{\cal Z}[g]
\ =\ \int_{0}^{-\infty} {\rmd g\over\pi}\,g^{-n-1}\,{\rm Im}({\cal
Z}[g+i0^+]) \ .
\label{intzn}
\ee
To obtain the behavior for large $n$, it turns out that it is
sufficient to evaluate the integral in (\ref{intzn}) in a saddle
point approximation.
Indeed, we shall show that, at least for $0<\epsilon<D$, the
integral is at large $n$ dominated by the discontinuity of ${\cal
Z}[g]$ at small negative $g$.
Moreover, ${\cal Z}[g]$ is dominated by a saddle
point, when re-expressed as a functional integral over properly
defined auxiliary fields.

The Hamiltonian (\ref{hamiltonian}) is non-local and involves a
distribution
of the field $\vec r$. It is convenient to introduce as an auxiliary
field the density of the membrane in bulk space
\be
\rho(\vec r)\ =\ \int \rmd^{D}\!x\ \delta^{d}(\vec r -\vec r(x))
\label{defrho}
\ee
and to write the interaction term as
\be
\int \rmd^{D}\!x\,\int \rmd^{D}\!y\,\delta^d(\vec r(x)-\vec r(y))
\ =\ \int \rmd^{d}\vec r\, \rho(\vec r)^2
\label{interrho}
\ .
\ee
Introducing a potential $V(\vec r)$ conjugate to $\rho(\vec r)$, we
can re-express
the interaction term in the partition function through a
Hubbard-Stratonovich transformation as
\begin{eqnarray}
\lefteqn{\exp\left[{-{b\over 2}\int \rmd^{D}\!x\,\int \rmd^{D}\!y\ \delta^d(\vec 
r(x)-\vec
r(y))}\right]}\nn\\
 &&= \int\! {\cal D}[\rho]\,\int\! {\cal
D}[V] \exp\left[{\,\int
\rmd^{d}\vec r\,\left(
V(r) \left[\rho(\vec r)-\int \rmd^{D}\!x\ \delta^{d}(\vec r -\vec
r(x))\right]
-{b\over 2}\,\rho(\vec r)^2\right)}\right]\hspace{5mm}
\label{Vandrho}
\end{eqnarray}
where
\be
\int {\cal D}[\rho]=\int_{-\infty}^{+\infty}\prod_{\vec
r\in{\mathbb{R}}^d}\rmd\rho(\vec r)
\ ,\qquad
\int{\cal D}[V]=\int_{-{\rm i}\infty}^{+{\rm i}\infty}\prod_{\vec r\in  
\mathbb{R}^d }
{\rmd V(\vec r)\over 2{\rm i}\pi}
\label{measrhoV}
\ .
\ee
Inserting \Eq{Vandrho} into \Eq{partition} and integrating over
$\rho$ yields
\be
{\cal Z}(b;L)\ =\ \int\!{\cal D}[\vec r]\int\!{\cal D}[V]
\,{\rm e}^{-{\cal H}[\vec r,V;b,L]}
\ee
with the new effective Hamiltonian
\be
{\cal H}'[\vec r,V;b,L]\ =\ {\int_L \rmd^{D}\!x\,\left({1\over
2}(\nabla\vec r(x))^2+V(\vec r(x))\right)\,-\,{1\over 2b}\,\int
\rmd^{d}\vec r\ V(\vec r)^2}
\label{HamrV}
\ .
\ee
This representation is nothing but the generalization of the well
known formulation of the Edwards model
as a model of free random walks in an (imaginary) annealed random
potential.
As above,  ${\cal Z}$ is a function  of the dimensionless
coupling $g$ and we
replace $b\to g$ and $L\to 1$ as in Eqs. \eq{dimlessg} and \eq{dimlessZ}.

As argued before, we aim in calculating the partition
function for small negative $g$.
For that purpose, it is convenient to rescale the coordinates and the
potential $V(\vec r)$
\be
x\,\to\,(-g)^{1\over D-\E}\, x
\quad,\quad
\vec r\,\to\,(-g)^{2-D\over2(D-\E)}\,\vec r
\quad,\quad
V\,\to\,(-g)^{-D\over D-\E}\,V \ ,
\label{brescaling}
\ee
so that we now consider a membrane with size $\bar L=(-g)^{-1\over
D-\E}$ and volume
\be
\bar{\cal V}\ =\ \bar L^D\ =\ (-g)^{-D\over D-\E} \ .
\label{voleff}
\ee
This yields the rescaled Hamiltonian
\begin{equation}
{\cal H}'_{\mathrm{resc}}[\vec r,V;{\bar L}]\ =\ {\int_{\bar L}
\rmd^{D}\!x\,\left({1\over
2}(\nabla\vec r(x))^2+V(\vec r(x))\right)\,-\,\frac{\bar{\cal
V}}2\,\int
\rmd^{d}\vec r\ V(\vec r)^2}
\label{HamrVresc}
\ .
\end{equation}
The  integral over $\vec r$ for fixed  potential $V$ defines
 the free energy {\it density} ${\cal E}[V]$ of a ``phantom''
(i.e.~non self-avoiding) membrane  in the external potential $V$
\begin{equation}
{\cal Z}[V]\ =\ {\rm e}^{-\,\bar{\cal V}\,{\cal E}[V]}
\ =\ \int\!{\cal D}[\vec r]\,
{\rm e}^{-\int_{\bar L}\rmd^{D}\!x\,{1\over 2}(\nabla\vec
r(x))^2\,+\,V(\vec r(x))} \ .
\label{freeen}
\end{equation}
The partition function finally becomes
\begin{equation}
{\cal Z}(g)\ =\ \int\!{\cal D}[V]\,{\rm e}^{-\,\bar{\cal
V}\left[{\cal E}[V]+{1\over 2}\int \rmd^{d}\vec r\,V(\vec r)^2\right]}
\label{Zfinal} \ \ .
\end{equation}
The crucial point of this formulation is that according to
 \Eq{voleff}, as long as
\begin{equation}
0\,<\,\epsilon\,<\,D \ ,
\label{epsleD}
\end{equation}
the limit $g\to 0^-$ corresponds to the thermodynamical limit when the
volume $\bar{\cal V}\to \infty$.
In this limit the free energy density ${\cal E}[V]$ 
has a finite limit (from extensivity) so that the volume appears only as a
global prefactor in the exponential of \eq{Zfinal}.
Hence in the large $\bar\mathcal{V}$ limit
the integral (\ref{Zfinal}) is dominated by a saddle point $V_\mathrm{inst}$,
which is an extremum of the effective energy ${\cal S}[V]$ for an 
\textit{infinite and flat} membrane.
The latter is defined as
\begin{equation}
{\cal S}[V]\,=\, {\cal E}[V]\,+\,{1\over 2}\,\int \rmd^{d}\vec r\,V(\vec
r)^2 \ ,
\label{actioninst}
\end{equation}
where ${\cal E}[V]$ is defined in (\ref{freeen}) as the free energy
density of an infinite flat  membrane in the potential $V$.
This saddle point $V_{\mathrm{inst}}(\vec r)$ is the non-trivial
instanton, since the action $\mathcal{S}$ of the trivial extremum $V(\vec r)=0$
is real and does
not contribute to the discontinuity of $\mathcal{Z}(g)$.
Moreover, as the instanton is obtained through the thermodynamical limit
$\bar L \to\infty$, it is independent of the shape of
the initial membrane.
This implies that the large order behavior of perturbation theory
is \textit{universal}, and does not depend on the internal geometry of the membrane. Let us now derive the saddle-point equations:

The variation of the free energy density is in general
\begin{equation}
{\delta{\cal E}[V]\over \delta V(\vec r)}
\ =\ 
\langle\delta[\vec r]\rangle_{V}
\label{varEwV}
\end{equation}
where $\delta[\vec r]$ is the normalized density of the membrane
\begin{equation}
	\delta[\vec r]\ =\ {\rho[\vec r]\over{\cal V}}\ =\ {1\over {\cal
V}}\int
	\rmd^{D}\!x\,\delta^{d}(\vec r-\vec r(x))
	\label{deltanorm}
\end{equation}
(which has a finite limit when the volume becomes infinite),
and ${\langle\ \rangle}_{V}$ denotes the expectation value for the
phantom membrane in the potential $V$, as defined in \Eq{freeen}.
Hence extremizing $\mathcal{S}[V]$ leads to the variational equation
for the instanton potential ${V_\mathrm{inst}}$
\begin{eqnarray}0\ &=&\ 
\langle\delta[\vec r]\rangle_{V_\mathrm{inst}} \ +\  
V_\mathrm{inst}(\vec r)
\label{equationV} \ .
\end{eqnarray}

Let us postpone the solution of \Eq{equationV} and first ask what
the
consequences of the existence of an instanton for the  large order
behavior are.
Denoting by ${\cal S}_{\mathrm{inst}}$ the action for the instanton
${\cal
S}[V_{\mathrm{inst}}]$,
we deduce from Eqs.~(\ref{voleff}) and (\ref{freeen}) that for small
negative $g$, the discontinuity of ${\cal
Z}(g)$ behaves as
\begin{equation}
{\rm Im}({\cal Z}(g))\ \approx\ \exp\left[{-(-g)^{-\,D\over
D-\epsilon}\,{\cal S}_{\mathrm{inst}}}\right]
\label{asympZ}
\end{equation}
and the integral representation for $z_n$ (\ref{intzn}) can be
evaluated by the saddle point method at large $n$.
This saddle point is at
\begin{equation}
g_{c}\ =\ -\,\left[{{\cal S}_{\mathrm{inst}}\over
n(1-\epsilon/D)}\right]^{1-\epsilon/D}
\label{gsaddle}
\end{equation}
and replacing the integral in (\ref{intzn}) by its value at $g_{c}$
gives the large $n$ behavior at leading order
\begin{equation}
z_n\ \sim\ \left(-\,{\cal C}\right)^n\,(n!)^{1-\epsilon/D}
\qquad,\qquad{\cal C}\ =\, \left[{1-\epsilon/D\over{\cal
S}_{\mathrm{inst}}}\right]^{1-\epsilon/D}
\label{largenz}
\end{equation}

Let us briefly discuss  this result.
For $0\le\E<D$, perturbation theory is divergent with
alternating signs.
For $\E=0$, one recovers the typical factorial behavior $(-{\cal C})^n
n!$ of field theories, provided that ${\cal S}_{\mathrm{inst}}$
remains UV finite.
As we shall see in the next section, our result (\ref{largenz})
coincides
for $D=1$ with the large order behavior deduced from the  $\Phi^4$
model with $n=0$ components.
The reasoning seems to break down  at $\E=D$, but we shall see
that
in fact the factor of ${\cal C}$, when considered as a function
of $D$ and
$\E$, is regular at $\E=D$ and can be continued to the
region
$\E\ge D$.
Thus, the asymptotics (\ref{largenz}), although derived for
$0<\E<D$, is valid in the whole physical domain $0<\E<2D$.
A more rigorous argument is as follows: Eqs.~(\ref{asympZ}) and
(\ref{gsaddle})  are still valid for $\epsilon>D$; the
instanton then governs the  behavior of the
discontinuity of $\mathcal{Z}(g)$ at \textit{large} $g$. This means
that the saddle point of \Eq{largenz}
for large $n$ now is at large negative $g$.

To go beyond this estimates, one must {\it (i)} compute the instanton
and its action, and
{\it (ii)} integrate the fluctuations around the instanton in
(\ref{Zfinal}).
If one assumes that this calculation goes along the same lines as
 in standard field theory, one must first isolate the zero modes,
i.e.\ the collective coordinates of the instanton.
As we shall see later, the instanton $V_{\mathrm{inst}}$
is rotationally invariant and
is
characterized
by its position in $d$-dimensional space only.
Thus it has $d$ zero modes, each of them gives a factor
$\mathcal{V}^{1/2}$ (by a standard collective coordinates argument),
and the remaining fluctuations
$\delta_{\perp} V$ (orthogonal to the translational variations
$\delta_{\mu} V\sim{\partial
V_{\mathrm{inst}}\over\partial r^{\mu}}$) give a finite determinant
${\cal
A}$.
Therefore we expect the semiclassical estimate for the
discontinuity to be
\begin{equation}
{\rm Im}({\cal Z}(g))\ \simeq\ {\cal A}^{-1/2}\,
\bar{\cal V}^{d\over 2}\,{\rm e}^{-\bar{\cal V}\,{\cal
S}_{\mathrm{inst}}}
\label{asymptZ}
\end{equation}
and that the large $n$ behavior is more precisely
\begin{equation}
z_n\ \sim\ {\cal A}'\ n^{d/2}\,\left(-\,{\cal C}
\right)^n\,(n!)^{1-\epsilon/D}\ \left[ 1\,+\,\ldots\right] \ .
\label{asympz}
\end{equation}

Finally we shall see that the action
of the instanton remains finite in the limit $\epsilon\to 0$.
As in standard $\Phi^4$ theory, one expects UV divergences
to appear
only for fluctuations around the instanton, and that
these divergences are cancelled by the same renormalizations as in
perturbation theory.
This would imply that our large order estimate (\ref{largenz})
is also valid  for the renormalized theory at $\epsilon=0$, in
particular for the renormalization group functions which enter into
the  $\epsilon$-expansion of the scaling exponents.
Renormalization however has to be taken into account when evaluating the
constant ${\cal A}'$ in (\ref{asympz}).

\section{The polymer case and physical interpretation of the
instanton}
\label{s.3}

Before discussing  membranes, let us study in
detail the special case $D=1$, where the model reduces to  the
Edwards model for polymers.
Using the well known mapping between the problem of a Brownian walk in a
potential $V(\vec r)$ and quantum mechanics of a single particle in
the same potential, the free energy density ${\cal E}[V]$ of a
linear chain fluctuating in a potential $V(\vec r)$  is in the
thermodynamic limit given by the  lowest eigenvalue $E_{0}$ of the operator
\begin{equation}
	H\ =\ -{\Delta\over 2}\,+\,V(\vec r)
	\label{Hoperator} \ ,
\end{equation}
where $\Delta$ is the Laplacian in $d$ dimensions.
Thus we have
\begin{equation}
	{\cal E}[V]\ =\ E_{0} \ .
	\label{EeqE0}
\end{equation}
Denoting by $\Psi_{0}(\vec r)$ the ground state wave-function, and
using Eq.~(\ref{equationV}) and the standard result from  first order
perturbation theory

\begin{equation}
\langle\delta[\vec r]\rangle \raisebox{-1.5mm}{$_V$}\ =\
{\delta{E_{0}}[V]\over \delta V(\vec r)}\ =\ \langle\Psi_{0}|{\delta H\over  
\delta V(\vec
r)}|\Psi_{0}\rangle\ =\ |\Psi_{0}(\vec r)|^{2}
	\label{varE0} \ ,
\end{equation}
we obtain the instanton potential
\begin{equation}
V_{\mathrm{inst}}(\vec r)\ =\ -\,\left(\Psi_{0}(\vec r)\right)^{2}
\ .
\label{VPsi}
\end{equation}
The eigenvalue equation $H\Psi_{0}=E_{0}\Psi_{0}$ becomes  non-linear
\begin{equation}
\,{1\over 2}\,\Delta\,\Psi_{0}\,+\,E_{0}\,\Psi_{0}\,+\,{\Psi_{0}}^{3}
\ =\ 0
 \ .
\label{eqins1D}
\end{equation}
Since $\Psi_{0}$ obeys the normalization condition
\begin{equation}
	\|\Psi_{0}\|^{2}\,=\,\int \rmd^{d}\vec r\,\Psi_{0}(\vec r)^{2}\,=\,1 \ ,
	\label{normpsi0}
\end{equation}
the wave function $\Psi_{0}$ and the ground state
energy $E_{0}$ are fully determined
by \Eqs{eqins1D} and (\ref{normpsi0}).
Eq.~(\ref{eqins1D}) has nontrivial normalizable solutions for
$2<d<4$ and $E_{0}<0$. In addition, the ground state
$\Psi_{0}$ is rotational symmetric, i.e.\ does not vanish at finite $\vec r$.
The action for the instanton (\ref{actioninst}) finally reads
\begin{equation}
{\cal S}_{\mathrm{inst}}\ =\ E_{0}\,+{1\over 2}\int \rmd^{d}\vec r\,
{\Psi_{0}}^{4} \ .
\label{Sins1D}
\end{equation}

To make contact to the instanton analysis in the
Landau-Ginsburg-Wilson (LGW) $\Phi^{4}$-theory with $n=0$ components,
remark that
Eqs.~(\ref{eqins1D})-(\ref{normpsi0}) hold \textit{if and only if} $\Psi_{0}$ 
and $E_{0}$ are extrema of the action
\begin{equation}
	\mathcal{S}'[\Psi,E]\ =\ E\,+\ \int \rmd^{d}\vec r \ \left[{1\over 2}
	(\nabla\Psi)^{2}\,-\,E\,\Psi^{2}\,-\ {1\over 2}\,\Psi^{4}\right]
	\label{actionEPsi}\ .
\end{equation}
This is the  standard Landau Ginsburg Wilson action with negative
coupling associated to $\Psi^4$ and mass $m^{2}=-2E$.
Moreover, \textit{at the extrema}, the two actions
are equal
\begin{equation}
	\mathcal{S}_{\mathrm{inst}}\left[\Psi_0,E_0 \right]\ =\  
\mathcal{S}'\left[\Psi_0,E_0 \right]\ .
	\label{SinsteqS}
\end{equation}
The relation becomes clearer by the change of variables
\begin{equation}
	\Psi(\vec r)\ =\ \left({-2E\over 4-d}\right)^{\half}\,
	\Phi\left({\textstyle\left( {-2E\over 4-d}\right)^{\half}}\vec r\right)
	\label{Psi2Phi} \ .
\end{equation}
The action $\mathcal{S}'$ then reads
\begin{equation}
	\mathcal{S}'[\Psi,E]\ =\ E\ +\ \left({-2E\over 4-d}\right)^{2-{d\over 2}}\,
	\mathcal{S}_{\mathrm{LGW}}[\Phi]
	\label{SEPhi}
\end{equation}
with
\begin{equation}
	\mathcal{S}_{\mathrm{LGW}}[\Phi]\ =\
	\int \rmd^{d}\vec r\ \left[{1\over 2}\,(\nabla\Phi)^{2}\,+\,
	{4-d\over 2}\,
	\Phi^{2}\,-\,{1\over 2}\,\Phi^{4}\right] \ .
	\label{SLGW}
\end{equation}
We can extremize (\ref{SEPhi}) with respect to $E$ and $\Phi$
independently, and denoting by $\Phi_{0}$ and $E_0$ these
extremizing solutions, we get
\be
E_0={\textstyle \left(\frac d2 -2 \right)}\, S_{\mathrm{LGW}}
\left[ \Phi_0\right]^{\frac1{d/2-1}} \ .
\label{ELGW}
\ee
The change of variables in \Eq{Psi2Phi} was constructed such that the
instanton action takes the simple form
\begin{equation}
	\mathcal{S}_{\mathrm{inst}}\ =\ \mathcal{S}'[\Psi_{0},E_{0}]\ =\
	{\textstyle \left(\frac d2 -1\right)}
	\,\mathcal{S}_{\mathrm{LGW}}[\Phi_{0}]^{{1\over d/2-1}}
	\label{SfromPhi} \ .
\end{equation}
Since for polymers ($D=1$) $d/2-1=1-\epsilon/D$, we can use
Eq.~(\ref{largenz}) to write the large order constant $\mathcal{C}$
of the Edwards model as
\begin{equation}
	\frac1{\mathcal{C}}\ =\ \mathcal{S}_{\mathrm{LGW}}[\Phi_{0}]
	\label{CfromPhi} \ .
\end{equation}
This result could have been derived directly from the standard field
theoretical formulation of the Edwards model as a $n=0$ component
${(\vec\Phi^2)}^2$ model.

\begin{figure}[t]
\centerline{
\parbox{4mm}{\raisebox{20mm}{$\displaystyle \frac 1 {\mathcal C}$}}
 \epsfxsize=0.7\textwidth  \parbox{0.7\textwidth}{
 \epsfbox{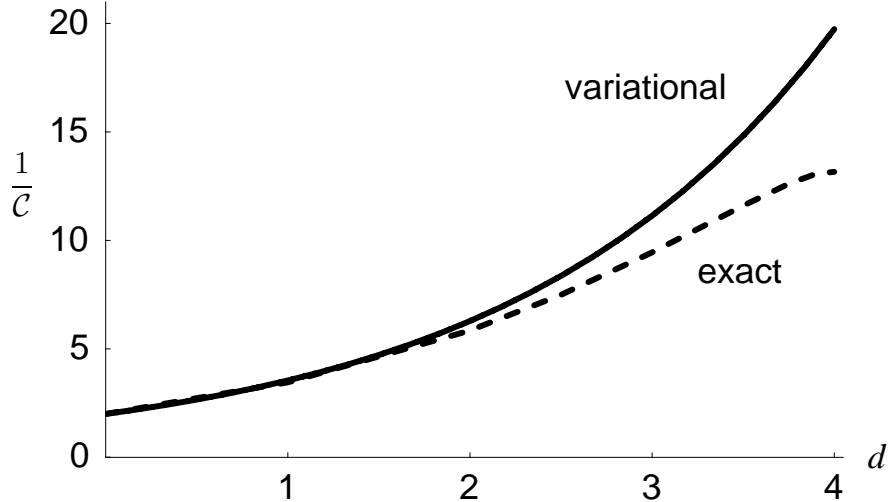}}}
 \caption{$1/\mathcal{C}$ as obtained from a   numerical solution of  
\protect\Eq{CfromPhi}, compared to the variational bound derived later in
\protect\Eq{Cinstvar}.}
\label{phi4-instanton-plot}
\end{figure}

The equation for the instanton derived from the action (\ref{SLGW})
admits a regular  solution $\Phi_{0}(|\vec r|)$ for any
$0\le d\le 4$, so that nothing special occurs at the point $d=2$
(i.e.\ $\epsilon=D=1$) as one might have expected from
Eq.~(\ref{largenz}).
Let us note that since the ``mass'' in \Eq{SLGW} is equal to $4-d$,
it is positive for $d<4$ but vanishes at the critical dimension $d=4$,
so that the instanton solution $\Phi_{0}$ still exists for $d=4$.
In Fig.~\ref{phi4-instanton-plot} we plot ${\cal C}^{-1}(d)$ for
$0\le d\le 4$, as obtained from numerical integration.

It is interesting to give a physical interpretation of the instanton for the  
Edwards model, since this interpretation is the same for membranes
with $D\ne 1$.
Let us first recall the standard interpretation of the instanton for the
LGW model with action (\ref{SLGW}), i.e.\ negative $\Phi^4$-coupling.
The classical false vacuum  $\Psi(r)=0$ is separated from the
true vacua  $\Psi(r)=\mp\infty$ by a finite barrier.
The instanton solution $\Psi_{0}$ describes a metastable droplet of true  
vacuum (with $\Psi_0(r)\ne 0$ inside the droplet) in the false vacuum, which  
is on the verge to nucleate. Indeed, if the droplet is slightly larger, the  
positive surface energy dominates and the droplet shrinks and finally  
vanishes,
 while if it is slightly smaller, the negative volume energy dominates and  
the droplet expands.

We now consider the energy density ${\cal S}[V]$ given by
\Eq{actioninst}.
It corresponds to the total free energy of a polymer globule trapped
in the potential well $V(\vec r)<0$, where this
effective potential results from the \textit{attractive} 2-point
interaction between elements of the polymer (since we are at negative
coupling, $b<0$).
To see how ${\cal S}$ varies with the average gyration radius of the
polymer, it is convenient to consider the following scale
transformation on $V$
\begin{equation}
V(\vec r)\ \to\ V_\lambda(\vec r)\ =\ \lambda^{2D\over
2-D}\,V(\lambda\vec r) \ .
\label{scaleV}
\end{equation}
Simple dimensional analysis shows that under (\ref{scaleV})
\begin{equation}
{\cal E}[V]\ \to\ \lambda^{2D\over 2-D}\,{\cal E}[V]\qquad,\qquad
\int V^2\ \to\ \lambda^{2\epsilon\over 2-D}\,\int V^2
\label{scaleE}
\end{equation}
(here $D=1$ and $\epsilon=2-d/2$).
As long as $\epsilon<D$, and for large $\lambda$, i.e.\ when shrinking the
polymer
globule, it is the first term ${\cal E}<0$ which dominates and the
total free energy ${\cal S}$ becomes large and negative; while for
small $\lambda$, i.e.\ when expanding the globule, it is the second
term on the r.h.s.\ of (\ref{actioninst}) which is larger than
$0$, and which
dominates.
Thus in this mean field picture, i.e.\ neglecting thermal
fluctuations around the instanton, large globules tend to expand, while small 
globules tend to collapse.
This has a simple physical interpretation: the polymer trapped in its
own potential is subject to two opposite forces, (i) attractive
forces between its elements which would like to make the polymer
collapse,
(ii) entropic
repulsion which exerts a pressure on the well and would like to
expand the polymer (until it becomes a free random walk).
What our calculation implies is the simple fact that for large radius
(i.e.\ small $\lambda$) entropic repulsion dominates, while at small
radius (large $\lambda$) attraction dominates and the polymer
collapses.
Thus the instanton solution describes a polymer with attractive
interactions on the verge to collapse into its dense (and most
stable) phase; this is similar to the instanton
in the LGW theory which describes a bubble of true vacuum on the
verge to
nucleate and to destroy the false vacuum.

\section{Gaussian variational calculation}
\label{s.4}
For $D\ne 1$ (and in general for $0<D<2$ non-integer) we know of no
exact method to calculate the instanton.
A simple and natural approximation is the variational method, i.e.
the Hartree-Fock approximation.

To evaluate the free energy density ${\cal E}[V]$ of the free,
 i.e.\ non-interacting membrane in
a potential $V$, and described by the Hamiltonian
\begin{equation}
{\cal H}_V\ =\ \int \rmd^{D}\!x\,\left({1\over 2} (\nabla\vec
r)^2\,+\,V(\vec r)\right)\ ,
\label{HamV}
\end{equation}
we introduce the trial Gaussian Hamiltonian
\begin{eqnarray}
{\cal H}_{\rm var} &=& \int \rmd^{D}\!x\int \rmd^D\!y\,{1\over 2}\,\vec  
r(x)\,K(x-y)\,\vec r(y)\nonumber\\
&=& \int \frac{\rmd^{D}k}{(2\pi)^{D}}\, {1\over 2}\,\vec{\tilde  
r}(k)\,\tilde K(k)\,\vec{\tilde r}(-k) \ ,
\label{Hamvar}
\end{eqnarray}
where ${\tilde{~}}$ denotes the Fourier transform.
The free energy for the trial Hamiltonian is
\begin{eqnarray}
{\cal E}_{\rm var}\ &=&\ -\,{1\over{\cal V}}\ {\rm Log}\left[\int
{\cal D}[\vec r]\,{\rm e}^{-{\cal H}_{\rm var}}\right]\nonumber\\
&=&\ {d\over 2}\int{\rmd^{D}k\over {(2\pi)}^{D}}\ {\rm Log}[\tilde
K(k)/k^2] \ ,
\label{Evar}
\end{eqnarray}
and the factor of $1/k^2$ comes from the normalization of the measure
${\cal
D}[\vec r]$ taken such  that ${\cal E}[V=0]=0$.
${\cal V} $ is  the total volume of the membrane.

The Hartree Fock approximation amounts in replacing ${\cal E}[V]$ by
the best variational estimate ${\cal E}_{\rm var}[V]$
\begin{equation}
{\cal E}[V]\,\le\ {\cal E}_{\rm var}[V]\ =\ {\cal E}_{\rm
var}\,+\,{1\over{\cal V}}\,\left.\langle{\cal H}_V-{\cal H}_{\rm
var}\rangle\right._{\rm var} \ .
\label{EVvar}
\end{equation}
$\langle\ \rangle_{\rm var}$ denotes the average with respect to the
trial Hamiltonian ${\cal H}_{\rm var}$ and
 one must look for the trial Hamiltonian ${\cal H}_{\rm var}$
(i.e.\ the kernel $K$) which  minimizes
${\cal E}_{\rm var}[V]$.
Denote by $\tilde V(\vec p)$ the Fourier transform
 of the potential $V(\vec r)$.
Since the variational Hamiltonian is Gaussian, it is easy to compute
the second term on the r.h.s.\ of (\ref{EVvar}),
${\cal V}^{-1}\langle{\cal H}_V-{\cal H}_{\rm var}\rangle_{\rm var}$
in the infinite volume limit:
\begin{eqnarray}
\lefteqn{\langle V(\vec r(0)\rangle_{\rm var}\
+\ {1\over 2}\,\left[
\langle\nabla\vec r(0))^2\rangle_{\rm var}
\,-\,\int \rmd^D\!x\,K(x)\,\langle\vec r(x)\vec r(0)\rangle_{\rm var}
\right]} \nonumber\\
&=&\,\int {{\rmd^d}\vec p\over (2\pi)^d}\,\tilde V(\vec p)\,\langle {\rm
e}^{{\rm i}\vec p \vec r(0)}\rangle_{\rm var}\ +\ \int {\rmd^D k\over
(2\pi)^D}\, {k^2-\tilde K(k)\over 2}\,\langle\vec{\tilde
r}(k)\vec{\tilde r}(-k)\rangle_{\rm var}
\nonumber\\
&=&\,
\int {{\rmd^d}\vec p\over (2\pi)^d}\,\tilde V(\vec p)\,
\exp\left[
{-{\vec p^2\over 2}\int {\rmd^D k\over (2\pi)^D}{1\over \tilde
K(k)}}
\right]
\,+\,{d\over 2}\int {\rmd^D k\over (2\pi)^D}\,\left({k^2\over\tilde
K(k)}-1\right) \ .
\label{2ndvar}
\end{eqnarray}
Combining \Eqs{actioninst}, (\ref{Evar}) and (\ref{2ndvar}), we
finally obtain the variational estimate for the total energy of the
instanton
\begin{eqnarray}
{\cal S}_{\rm var}[V]\ &=&\ {\cal E}_{\rm var}[V]\,+\,{1\over 2}\int
\rmd^d\vec r\,
V(\vec r)^2\nonumber\\
&=&\
\int {{\rmd^d}\vec p\over (2\pi)^d}\,\left(\tilde V(\vec p)\,
\exp\left( -{\vec p^2\over 2}\int {\rmd^D k\over (2\pi)^D}{1\over \tilde
K(k)}\right)
 +\ {1\over 2} \tilde V(\vec p)\tilde V(-\vec p)\right)
\nonumber\\
& &\ +\ {d\over 2}\int {\rmd^D k\over (2\pi)^D}\,\left({\rm
Log}\left[\frac{\tilde K(k)}{k^2}\right]\,+\,{k^2\over\tilde K(k)}-1\right)
 \ .
\label{Svarfinal}
\end{eqnarray}
We now extremize \Eq{Svarfinal} {\it both} with respect to $\tilde
K$
(variational approximation) and with respect to $\tilde V$ (to
obtain the instanton solution).
Extremizing w.r.t.\ $\tilde K(k)$ yields the equation
\begin{equation}
\tilde K(k)\ =\ k^2\ -\ {1\over d}\int {\rmd^d \vec p\over
(2\pi)^d}\,{\vec p}^2\,
\tilde V(\vec p)\,\exp\left[{-{\vec p^2\over 2}\int {\rmd^D k\over
(2\pi)^D}{1\over \tilde K(k)}}\right]
\label{equvarK}
\end{equation}
which implies that the variational Hamiltonian depends just on a mass
$m_{\rm var}$
\begin{equation}
\tilde K(k)\ =\ k^2\,+\,m_{\rm var}^2 \ .
\label{Kform}
\end{equation}
Extremizing \Eq{Svarfinal} w.r.t.\ $\tilde V(p)$ gives
\begin{equation}
\tilde V_{\mathrm{inst}}^{\rm var}(\vec p)\ =\
-\ \exp\left[{-{\vec p^2\over 2}\,A}\right]
\label{equvarV}
\end{equation}
with
\begin{equation}
A\ =\ \int {\rmd^D k\over (2\pi)^D}{1\over \tilde K(k)}\ =
 m_{\rm var}^{D-2}\,{\Gamma(1-{D\over
2})\over(4\pi)^{D/2}} \ .
\label{equA}
\end{equation}
$\Gamma$ is Euler's Gamma function.
Thus, in the variational approximation the instanton potential is
Gaussian.
Inserting \Eq{equvarV} into \Eq{equvarK} yields the
self-consistent equation for $m_{\rm var}$
\begin{equation}
m_{\rm var}^{2}\ =\ {1\over d}\int{\rmd^d\vec p\over (2\pi)^d}\,\vec
p^2\,{\rm e}^{-\vec p^2\,A}\ =\ {1\over 2}\,(4\pi)^{-d/2}\,A^{-1-d/2}
\label{eqmvar} \ .
\end{equation}
We finally get in terms of $D$, $\E$ and $d=2(2D-\epsilon)/(2-D)$
\begin{equation}
m_{\mathrm{var}}\ =\
\sqrt{4\pi}\,\left[2\,\Gamma({\scriptstyle{2-D\over 2}})^{1+{d\over
2}}\right]^{1\over D-\epsilon} \ .
\label{mvarfinal}
\end{equation}
The final result for $A$ reads 
\be
A=\frac1{4\pi} \Gamma\left(\textstyle\frac{2-D}2\right)^{\frac{-2}{D-\E}}
2^{\frac{D-2}{D-\E}} \ \ .
\ee
We can now insert these results into  \Eq{Svarfinal}, and after
straightforward calculations get the variational instanton action
\begin{equation}
S^{\rm var}_{\mathrm{inst}}\ =\ {\cal S}_{\rm
var}[V_{\mathrm{inst}}^{\rm var}]\ =\
\left(1-{\epsilon\over
D}\right)\,\left[2\,\Gamma({\scriptstyle{2-D\over 2}})
^{d\over D}\right]^{D\over D-\epsilon}
\label{Sinstvar} \ .
\end{equation}
The corresponding variational estimate for the large order
constant ${\cal C}$ defined by Eq.~(\ref{largenz}) is
\begin{equation}
1/{\cal C}^{\rm var}\ =\ 2\ \Gamma\!\left({\scriptstyle{2-D\over
2}}\right)^{d\over D}
\ .
\label{Cinstvar}
\end{equation}
As claimed in the previous section, although intermediate results are
singular at $\epsilon=D$, the final result is regular for all
$\epsilon>0$.
We shall discuss the physical significance of these results in the
next section.

\section{Discussion of the variational result}
\label{s.5}
\subsection{$D=1$}\label{s.5.1}
It is interesting to compare the variational estimate with the
exact result for polymers, i.e.\ for the case $D=1$.
Let us consider the LGW instanton action, as given by
Eq.~(\ref{SLGW}).
It is equal to the inverse of the large order constant  $\mathcal{C}$.
On \Fig{phi4-instanton-plot} we have plotted the variational result for
$1/\mathcal{C}^{\mathrm{var}}$, as given by \Eq{Cinstvar} and
the exact result for $1/\mathcal{C}$ obtained by numerical solution,
as a function of $0<d<4$.
First we note that  always
\begin{equation}
	\mathcal{C}\ \ge\ \mathcal{C}^{\mathrm{var}}
	\label{inequC}
\end{equation}
as expected from the variational inequality
$\mathcal{E}\le\mathcal{E}_{\mathrm{var}}$.
This implies that the variational method gives an underestimate of
the large orders.

Second the variational estimate becomes good for small $d$, and
exact for $d\to 0$. This is not unexpected, since in that
limit the membrane $\mathcal{M}$ has no inner degrees of freedom, and the
functional integration over $V(r)$ reduces to a simple
integration over $V\in \mathbb{R}$. Since this integral is Gaussian,
the variational method becomes exact.

Finally, the variational estimate for $\mathcal{C}$ is regular when $d\to 4$,
and then equals  $1/(2\pi^{2})$; this
is  $50\% $ smaller than the exact result $3/(4\pi^{2})$.
Thus the variational method is only qualitatively correct when
$\epsilon=0$.
This is not so surprising, since the limit $\epsilon\to 0$ is somewhat peculiar.
Indeed when $d=4$ the ground state energy $E_0$ in the equation \eq{eqins1D} 
for the wave function $\Psi_0$ is then equal to $0$.
Then the most general solution to \Eq{eqins1D} (for $d=4$ and $E_0=0$) is 
\begin{equation}
\Psi_0(\vec r)\ =\ {2 r_0 \over r_0^2+\vec r^{\,2}}
\ ,
\label{inst4d}
\end{equation}
with $r_0$ an arbitrary scale (the size of the instanton).
$r_0$ is fixed by the normalization condition \eq{normpsi0} which cannot
be fulfilled at $d=4$ for finite $r_0$.
In fact a more careful analysis of the rotationally invariant solutions of
Eqs.~\eq{eqins1D} and \eq{normpsi0} (see Appendix~\ref{a.2})
shows that as $d\to 4$, $E_0$
should scale as 
$E_0\sim 4-d$ and that for $0<4-d\ll 1$ the true solution $\Psi_0$ is
well approximated by \Eq{inst4d} (at least as long as $|\vec r |^2 (4-d)\ll 1$)
with an instanton size $r_0$ which vanishes as $d\to 4$ as
\begin{equation}
r_0\ \sim\ {1\over\sqrt{|\log(4-d)|}}\ .
\label{r0dto4}
\end{equation}
The corresponding instanton potential $V_{\mathrm{inst}}=-|\Psi_0|^2$ is
also singular in the $d\to 4$ limit (it may be considered as a Dirac-like $\delta$-function), and is very poorly approximated by the Gaussian
variational solution at $D=1$, $d=4$ for the potential
\be
V^{\mathrm{var}}\ =\ -\,16 \pi^2 \rme^{-16 \pi^2 \vec r^{\,2}} \  ,
\label{varD1d4}
\ee
with positive width.
As usual with variational methods, the approximation for the ground state
energy is much better than that for the wave function.

\subsection{Consequences for the $\epsilon$-expansion}
Of course, one is interested in the consequences of these large order
estimates for the $\epsilon$-expansion of the scaling
exponents for self-avoiding membranes and polymers.
Let us recall that in renormalized perturbation theory
one computes the renormalization group $\beta$-function $\beta(g)$, 
as a power series in $g$ of the form
\footnote{Strictly speaking $g$ is now the renormalized coupling constant $g_{\mathrm{R}}$.}
\begin{equation}
\beta(g)\ =\ -\ \epsilon\,g \ +\ B_{1}\,g^{2}\ +\ \mathcal{O}(g^{3})
\ .
\label{Wfunction}
\end{equation}
Its zero at $g^{*}=\epsilon/B_{1}+\mathcal{O}(\epsilon^2)$ is the
IR fixed point which governs the scaling limit for large membranes.
Other anomalous dimensions, like the dimension $\nu(g)$
of the field $\vec r$
(which gives the fractal dimension of the membrane) can also
be computed as a series in $g$.
Their values at the fixed point $g^{*}$ give the scaling exponents
of the membrane, and may be expanded as power series in $\epsilon$.

By analogy with the ordinary Wilson-Fisher $\varepsilon$-expansion
for LGW field theories, let us assume that the large orders of the
function $\beta(g)$ and of the other anomalous dimensions are
given by the instanton estimate, and that
they can be resummed by Borel-techniques.
We are not able at the moment to give any more precise argument to
this last claim (which is still a conjecture even for the LGW theories).
Then a simple calculation consists in estimating the ``optimal''
order $n_{\mathrm{opt}}$ beyond which the $\epsilon$-expansion starts
to diverge.
If we only know the first $n$ terms of the expansion,
we expect that for $n<n_{\mathrm{opt}}$ ``ordinary'' resummation
procedures (like Pad\'e) will be sufficient.
If $n>n_{\mathrm{opt}}$, or if one seeks higher precision,
knowledge of the large orders  and
more sophisticated resummation methods are required.
Assuming that for $\epsilon=0$ the $n$-th coefficient
 of $\beta(g)$ is of order $(-\mathcal{C}g)^{n}n!$, and
that we can approximate the fixed point $g^{*}$ by its first order
estimate $\epsilon/B_{1}$, the term
of order $n$ in the $\epsilon$ expansion  should behave as
\begin{equation}
	\left(-\,{\epsilon\ \mathcal{C}\over B_{1}}\right)^{n}\ n!
	\label{largeneps} \ \ 
\end{equation}
The optimal order $n_{\mathrm{opt}}$ is obtained when the
absolute value of (\ref{largeneps}) is the smallest, that is for
\begin{equation}
	n_{\mathrm{opt}}\ \epsilon\ \simeq\ 
{B_{1}\over\mathcal{C}} \ ,
	\label{nopteps}
\end{equation}
where $B_{1}$ is the one-loop coefficient of the $\beta$-%
function, and $\mathcal{C}$ the large order constant as obtained from
the instanton calculus.

\begin{figure}[t]
\unitlength=1.cm
\begin{center}
\begin{picture}(11,7.5)
\put(1,1){\includegraphics{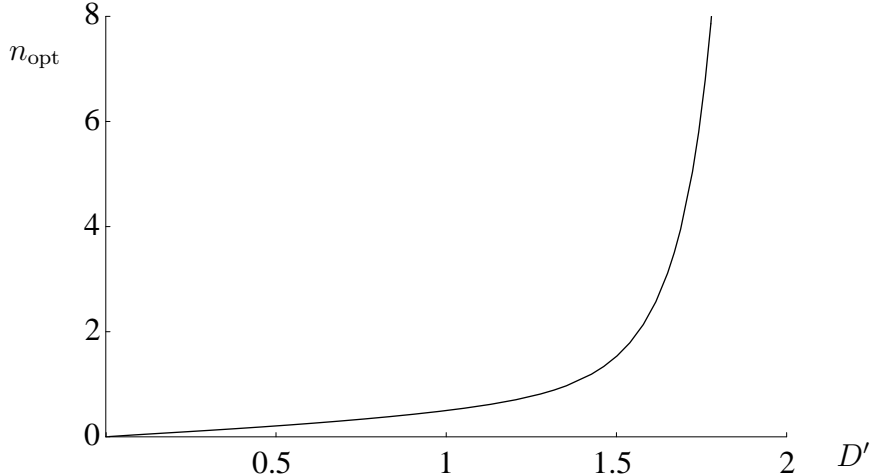}}
\put(11,1.1){$D'$}
\put(0,6.5){$n_{\mathrm{opt}}$}
\end{picture}
\end{center}
\vskip-7.ex
\caption{Optimal order $n_{\mathrm{opt}}(D')$ for the
$\epsilon$-expansion for membrane as function of the extrapolation
(dimension) parameter $D'$, as obtained from the variational estimate
for the large orders.}
\label{FnoptD}
\end{figure}

With our choice of normalizations for the coupling constant $b$ in the
Hamiltonian (\ref{hamiltonian}), the 1-loop coefficient of the
$\beta$-function is
\begin{equation}
B_{1}\ =\ \half \left[{(2-D)S_{D}\over 4\pi}\right]^{d/2}\, S_{D}^{2}\,
\left[1+{1\over 2-D}{
\left(\Gamma\left({\scriptstyle D\over 2-D}\right)\right)^{2}
\over \Gamma\left({\scriptstyle 2D\over 2-D}\right)}\right] \ ,
\label{W1expl}
\end{equation}
with $S_{D}$ the volume of the unit sphere in $D$ dimensions
\begin{equation}
S_{D}\ =\ {2\,\pi^{D/2}\over \Gamma\left({\scriptstyle D\over
2}\right)} \ .
\label{volsphere}
\end{equation}
Let us replace $\mathcal{C}$ in \Eq{nopteps} by the variational
approximant
$\mathcal{C}^{\mathrm{var}}$ given by Eq.~(\ref{Cinstvar}).
Setting finally $\epsilon=0$ in $B_1/\mathcal{C}$ (since we are
interested in the expansion around $\epsilon=0$), we obtain the
following variational estimate
for the r.h.s.\ of (\ref{nopteps})
\begin{equation}
 n_{\mathrm{opt}}\ \epsilon\ \simeq\
\frac{16}{(2-D)^2}
 \left[
 \Gamma\left({4-D\over 2}\right)
 \over
 \Gamma\left({D\over 2}\right)
 \right]^{{4\over 2-D}}\,
 \left[1+{1\over 2-D}{
 \left(\Gamma\left({\scriptstyle D\over 2-D}\right)\right)^{2}
 \over \Gamma\left({\scriptstyle 2D\over 2-D}\right)}
 \right]
\label{noptvar}
\end{equation}
Let us recall that in practice the $\epsilon$-expansion is used as
follows:
in order to compute for instance the scaling exponent $\nu$
for a membrane with internal dimension $D=2$
in $d$-dimensional space, one starts from some point $D'\ne D$,
$\epsilon=0$
(i.e.\ $d'=4D'/(2-D')$), uses an expansion in $\epsilon$ and $D-D'$
(or some more general expansion parameters) to evaluate
$\nu(D')$, which thus depends on the expansion point $D'$.
$\nu$ is then taken as the best estimate $\nu(D'_{\mathrm{opt}})$,
as determined for instance by a minimal sensitivity criterium.
Membranes ($D=2$)
 always correspond to $\epsilon=4$, so setting $\epsilon=4$ and
replacing $D$ by $D'$ in Eq.~(\ref{noptvar}), should give an
estimate of the ``optimal order'' $n_{\mathrm{opt}}(D')$ for the
$\epsilon$-expansion at $D'$.
The result for $n_{\mathrm{opt}}(D')$ is plotted on Fig.~\ref{FnoptD}.

Some interesting comments can be made on this curve.
For $D'>1.6$, $n_{\mathrm{opt}}(D')>2$ and becomes large as $D'\to 2$,
while for $D'<1.6$, $n_{\mathrm{opt}}(D')<2$ and becomes small as
$D'\to 0$.
In the first regime ($D'\to 2$) we thus expect that the power series
in $\epsilon$ will behave like a convergent series, up to some quite
large order $n_{\mathrm{opt}}$.
In the second regime ($D'$ small), we expect that the power series in
$\epsilon$ will be divergent from the very first terms.
This is in agreement with the  calculations at second order in 
\cite{DavWie96,WieDav97}.
For large $d$,  the 2-loop results for $\nu$ can neatly be resummed,
and the stability of the  various resummation procedures
and extrapolation schemes analyzed in 
\cite{DavWie96,WieDav97}
is good.
The final estimates are close to the prediction of a variational
approximation $4/d$ for $\nu$.
For smaller values of $d$ stability is less good, but in all cases,
the reliable extrapolations are obtained for values of the
extrapolation dimension $D'\simeq 1.6$ or  larger.
It is not possible to resum safely the 2-loop results if one starts
the $\epsilon$-expansions from $D'\le 1.5$.
Thus it seems that our rough estimates for the large order behavior may
explain some
general features of the calculation at second order, and
corroborate the results of the estimates of 
\cite{DavWie96,WieDav97}.

\subsection{Limit $D\to 2$}
Of course these arguments are valid if the variational approximation
for the
instanton action stays (at least qualitatively) correct in the limit
$D\to 2$.
First let us note that, although \Eqs{Sinstvar} and \eq{Cinstvar}
give estimates for $\mathcal{S}_{\mathrm{inst}}$ and $\mathcal{C}$
which are
singular  when $D\to 2$, our variational formula for
$n_{\mathrm{opt}}$
is much less singular, since according to  \Eq{noptvar} it behaves as%
\begin{equation}
n_{\mathrm{opt}}(D)\ \simeq\ {1\over\epsilon}\,
{16\,\mathrm{e}^{-4\gamma}\over (2-D)^2}
\qquad\mbox{as }D\to 2\ ,\ \epsilon\ \mbox{fixed} \ ,
\label{noptdto2}
\end{equation}
with $\gamma=0.577216$ the Euler's constant.
It has also been noted in \cite{Dup87} that instead of using the simple
coupling constant $b$ in the Hamiltonian (\ref{hamiltonian}) it might
be more physical and convenient to use as coupling constant the
``second virial coefficient"
$z$, defined as
\begin{equation}
z\ =\ \left[{(2-D)S_D\over 4\pi}\right]^{d/2}\, b\,L^{\epsilon} \ .
\label{zvirial}
\end{equation}
Using $z$ as expansion parameter instead of $g=bL^{\epsilon}$,
the large order constant $\mathcal{C}$ in
\Eqs{defzn} and \eq{largenz} is
now
\begin{equation}
\mathcal{C}_z\ =\ \mathcal{C}\,\left[{(2-D)S_D\over
4\pi}\right]^{-d/2} \ ,
\label{Cz}
\end{equation}
which in the variational approximation reads
\begin{equation}
\mathcal{C}_z^{\mathrm{var}}\ =\ \half
\left[{(2-D)S_D\over 4\pi}\Gamma\!\left(\scriptstyle{2-D\over
2}\right)^{2\over D}\right]^{-{d\over 2}}
\!\simeq\ \mathrm{Cst}\,\left({2-D\over 2}\right)^{2-{\epsilon\over
2}}\quad\mbox{as}\ D\to 2\, ,\ \epsilon\ \mbox{fixed} \ .
\label{Czvar}
\end{equation}
(with $\mathrm{Cst}=\half(\pi\,\mathrm{e}^{-2\gamma})^{2-\epsilon/2}$.
Therefore in this normalization also the singularities as $D\to 2$
are simply algebraic.
The same remark holds for the coupling constant normalization used in
\cite{DavWie96,WieDav97}.

This prompts us to study the consistency of the variational
approximation
in the limit $D\to 2$ (or equivalently $d\to\infty$) for fixed
$\epsilon$.
As we shall see, in fact the variational approximation becomes exact
in that
limit.
This makes the arguments of this section fully valid.

\section{Beyond the variational approximation and ${1/d}$ corrections}
\label{s.6}
\newcommand{\rmi}{\mathrm{i}}
\newcommand{\hvar}{\mathcal{H}_{\mathrm{var}}}
\newcommand{\mvar}{m_{\mathrm{var}}}
\newcommand{\Cvar}{\mathbb{C}_{\mathrm{var}}}
\newcommand{\Cc}{\mathbb{C}}
\newcommand{\Vvar}{V_{\mathrm{var}}}
\newcommand{\GA}{\includegraphics[scale=1]{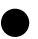}}
\newcommand{\GB}{\includegraphics[scale=1]{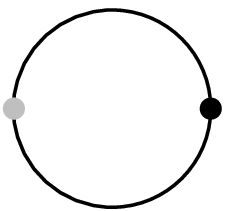}}
\newcommand{\GC}{\includegraphics[scale=1]{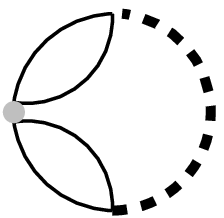}}
\newcommand{\GD}{\includegraphics[scale=1]{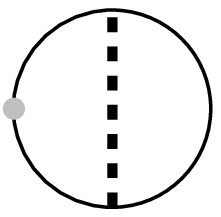}}
\newcommand{\GE}{\includegraphics[scale=1]{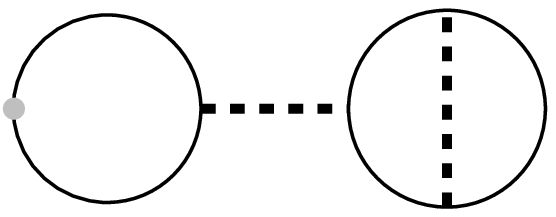}}
\newcommand{\GF}{\includegraphics[scale=1]{graph5.eps}}
\newcommand{\GG}{\includegraphics[scale=1]{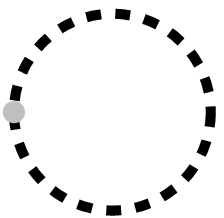}}
\newcommand{\GH}{\includegraphics[scale=1]{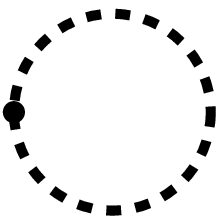}}
\newcommand{\GI}{\includegraphics[scale=1]{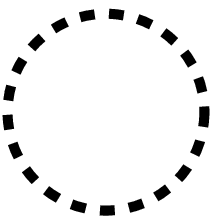}}
\newcommand{\GJ}{\includegraphics[scale=1]{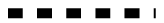}}
\newcommand{\GK}{\includegraphics[scale=1]{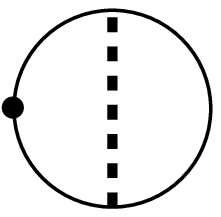}}
\newcommand{\GL}{\includegraphics[scale=1]{graph11.eps}}
\newcommand{\GM}{\includegraphics[scale=1]{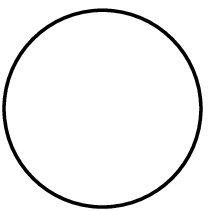}}
\newcommand{\GN}{\includegraphics[scale=1]{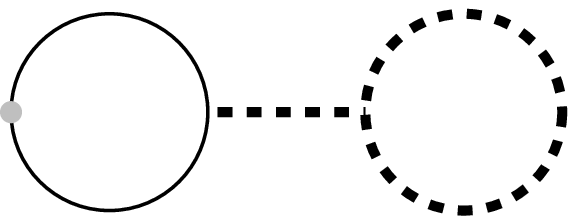}}
\newcommand{\GO}{\includegraphics[scale=1]{graph14.eps}}
\newcommand{\GP}{\includegraphics[scale=1]{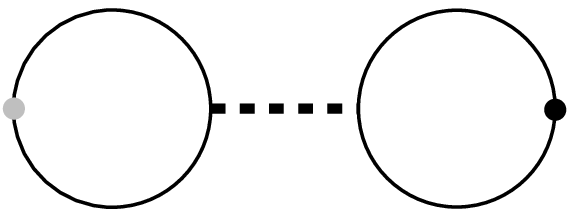}}
\newcommand{\GQ}{\includegraphics[scale=1]{graph16.eps}}
\newcommand{\Ga}{\raisebox{-.1cm}{\unitlength=1cm
\begin{picture}(.2,0.25)\put(0,0){\GA}\end{picture}}}
\newcommand{\Gb}{\raisebox{-1cm}{\unitlength=1cm
\begin{picture}(2.2,2.1)\put(0,0){\GB}\end{picture}}}
\newcommand{\Gc}[1]{\raisebox{-1cm}{\unitlength=1cm
\begin{picture}(2,2.1)\put(0,0){\GC}\put(1.4,1){\fbox{#1}}\end{picture}}}
\newcommand{\Gd}[1]{\raisebox{-1cm}{\unitlength=1cm
\begin{picture}(2.2,2.1)\put(0,0){\GD}\put(1.25,1){\fbox{#1}}\end{picture}}}
\newcommand{\Ge}[2]{\raisebox{-1cm}{\unitlength=1cm
\begin{picture}(5.5,2.1)\put(0,0){\GE}\put(2.6,.5){\fbox{#1}}
\put(4.75,1){\fbox{#2}}\end{picture}}}
\newcommand{\Gf}[2]{\raisebox{-1cm}{\unitlength=1cm
\begin{picture}(3.5,2.1)\put(0,0){\GF}\put(.7,.5){\fbox{#1}}
\put(2.85,1){\fbox{#2}}\end{picture}}}
\newcommand{\Gg}[1]{\raisebox{-1cm}{\unitlength=1cm
\begin{picture}(2.2,2.1)\put(0,0){\GG}\put(1.4,1){\fbox{#1}}\end{picture}}}
\newcommand{\Gh}[1]{\raisebox{-1cm}{\unitlength=1cm
\begin{picture}(2.2,2.1)\put(0,0){\GH}\put(1.4,1){\fbox{#1}}\end{picture}}}
\newcommand{\Gi}[1]{\raisebox{-1cm}{\unitlength=1cm
\begin{picture}(2.2,2.1)\put(0,0){\GI}\put(1.4,1){\fbox{#1}}\end{picture}}}
\newcommand{\Gj}[1]{\raisebox{-.5cm}{\unitlength=1cm
\begin{picture}(1.6,.5)\put(0,.5){\GJ}\put(.6,0){\fbox{#1}}\end{picture}}}
\newcommand{\Gk}[1]{\raisebox{-1cm}{\unitlength=1cm
\begin{picture}(2.2,2.1)\put(0,0){\GK}\put(1.25,1){\fbox{#1}}\end{picture}}}
\newcommand{\Gl}{\raisebox{-1cm}{\unitlength=1cm
\begin{picture}(2.2,2.1)\put(0,0){\GL}\end{picture}}}
\newcommand{\Gm}{\raisebox{-1cm}{\unitlength=1cm
\begin{picture}(2.2,2.1)\put(0,0){\GM}\end{picture}}}
\newcommand{\Gn}[2]{\raisebox{-1cm}{\unitlength=1cm
\begin{picture}(5.5,2.1)\put(0,0){\GN}\put(2.65,.5){\fbox{#1}}
\put(5.05,1){\fbox{#2}}\end{picture}}}
\newcommand{\Go}[2]{\raisebox{-1cm}{\unitlength=1cm
\begin{picture}(3.5,2.1)\put(0,0){\GO}\put(.75,.5){\fbox{#1}}
\put(3.1,1){\fbox{#2}}\end{picture}}}
\newcommand{\Gp}[1]{\raisebox{-1cm}{\unitlength=1cm
\begin{picture}(5.8,2.1)\put(0,0){\GP}\put(2.65,.5){\fbox{#1}}
\end{picture}}}
\newcommand{\Gq}[1]{\raisebox{-1cm}{\unitlength=1cm
\begin{picture}(4.,2.1)\put(0,0){\GQ}\put(2.65,.5){\fbox{#1}}
\end{picture}}}

In this section we show that the variational result is nothing but the leading term of a systematic expansion in $1/d$.
In ordinary theories like the one described by the effective Hamiltonian 
$\mathcal{H}_{V}$ 
which appears in \Eq{HamV} for a \textit{fixed potential} $V(\vec r)$ 
\begin{equation}
\mathcal{H}_{V}[\vec r]\ =\ \int \rmd^D\!x\,\left[
{1\over 2}(\nabla\vec r)^{2}+V(\vec r)\right] \ ,
\label{HamV2}
\end{equation}
this result is not unexpected.
For fixed $D$, the limit $d\to\infty$ is nothing but the limit where the
number of components $n=d$ of the field $\vec r$ becomes large.
In this limit, and provided that the model is O($n$) invariant,
it is known that the variational approximation becomes exact
and a systematic $1/n$ expansion can be constructed.
For the problem considered here, the existence of a $1/d$ expansion is not 
that easy to prove for two (related) reasons.
First, the potential $V(\vec r)$ is not fixed, it is also a variable which has
to be determined self-consistently, and it has a singular behavior when
$d\to\infty$.
In particular one cannot simply take the limit $d\to\infty$ while $D<2$ is
fixed, as can be seen on the exactly solvable case of polymers ($D=1$)
discussed in section~4.
Indeed in this case, the equation for the instanton has no physical
solution for $d>4$.
Second, the physically meaningful limit is to take $d\to\infty$ 
while $\epsilon=2D-d(2-D)/2$ is kept fixed.
In this limit, $D\to 2$ and one expects potentially dangerous additional
singularities, since the massive free propagator is known to 
have a logarithmic singularity at short distance for $D=2$.
There is a subtle interplay between the corresponding $1/(2-D)$ poles and
the terms which would naively disappear in the large $d$ limit.

\subsection{Expansion around the variational solution}

To evaluate the corrections to the variational approximation, we 
expand around the variational Hamiltonian
\begin{equation}
\mathcal{H}_{\mathrm{var}}[\vec r]\ =\ \int \rmd^D\!x\,\left[
{1\over 2}\,(\nabla\vec r)^{2}\,+\,{m^{2}_{\mathrm{var}}\over 2} \,\vec r^{\,2}
\right]
\label{HamVAr2}
\end{equation}
by writing
\begin{eqnarray}
\mathcal{H}_{V}[\vec r]\ &=&\ \mathcal{H}_{\mathrm{var}}[\vec r]
\,-\,\int\rmd^D\!x\ \Delta(\vec r)
\label{HV-Hvar}\\
\Delta(\vec r)\ &=&\ {m^{2}_{\mathrm{var}}\over 2} \,\vec r^{\,2}
\,-\,V(\vec r) \ .
\label{Deltadef}
\end{eqnarray}
The saddle point equation \eq{equationV} which defines the instanton potential 
$V$ is 
\begin{equation}
V(\vec r_0)\,+\,\langle\delta(\vec r_0-\vec r(x_0))
\rangle_{V}\ =\ 0 \ ,
\label{equV2}
\end{equation}
where $\langle\cdots\rangle_{V}$ denotes the expectation value of $\cdots$
taken with respect to the Hamiltonian $\mathcal{H}_{V}$
\begin{equation}
\langle\cdots\rangle_{V}\ =\ 
{\int\mathcal{D}[\vec r]\ \ldots\ \rme^{-\mathcal{H}_V}
\over
\int\mathcal{D}[\vec r]\ \rme^{-\mathcal{H}_V}} 
\ ,\label{expvalV}
\end{equation}
for an infinite flat membrane. (Recall that this was the limit we had
to take in \Eq{Zfinal}).
This implies that the point $x_0$ can be chosen arbitrarily on the membrane.
It is simpler to use the Fourier transform of $V$
\begin{equation}
\tilde V(\vec k)\ =\ \int \rmd^d\vec r\ \rme^{\mathrm{i}\vec k\vec r}\,V(\vec r)
\ ,\label{hatV}
\end{equation}
so that \Eq{equV2} reads
\begin{equation}
\tilde V(\vec k_0)\ +\ \langle\rme^{\mathrm{i}\vec k_0\vec r(x_0)}\rangle_{V}\ = 0 \ .
\label{equVhat}
\end{equation}
We now expand around the variational solution, using \Eq{HV-Hvar},
to rewrite the expectation value on the l.h.s.\ of \Eq{equVhat}
as a \textit{connected} correlation function computed with the variational
Hamiltonian
\begin{equation}
\langle\rme^{\rmi\vec k_0\vec r(x_0)}\rangle_V\ =\ 
\langle\rme^{\rmi\vec k_0\vec r(x_0)}\cdot\rme^{\int_x\!\Delta}
\rangle^{\mathbf{C}}_{\mathrm{var}}
\ ,\label{V2var}
\end{equation}
where $\langle\cdots\rangle_\mathrm{var}$ denotes the expectation value of
$\cdots$
taken with respect to the variational Hamiltonian $\mathcal{H}_{\mathrm{var}}$
\begin{equation}
\langle\cdots\rangle_{\mathrm{var}}\ =\ 
{\int\mathcal{D}[\vec r]\ \ldots\ \rme^{-\mathcal{H}_{\mathrm{var}}}
\over
\int\mathcal{D}[\vec r]\ \rme^{-\mathcal{H}_{\mathrm{var}}}} 
\ .\label{expvalvar}
\end{equation}
The suffix $\langle\cdots\rangle^{\mathbf{C}}$ means the \textbf{connected} correlation function in the usual sense:
since $\hvar$ is a free Gaussian Hamiltonian, using Wick's theorem, correlation
functions like that in \Eq{V2var} can be expressed as Feynman diagrams involving the
free variational propagator in $D$ dimensions 
\begin{equation}
G_{\!\mathrm{var}}(x)\ =\ 
\int {\rmd^Dq\over (2\pi)^D}\,{\rme^{-\rmi q\cdot x}\over
q^2+m_{\mathrm{var}}^2}
\label{propvar}
\end{equation}
and vertices obtained by expanding the ``perturbation term'' $\Delta(\vec r)$
of \Eq{Deltadef} in powers of $\vec r$;
the connected correlation function in \Eq{V2var} is just given by the restriction to connected diagrams. 

Similarly, the free energy density $\mathcal{E}[V]$ defined by \Eq{freeen}
can be written as a sum over connected vacuum diagrams.

\subsection{Resummation of tadpoles and reorganization in terms of normal products}
In these calculations, we encounter numerous ``tadpole diagrams'', which
result from the evaluation of $\langle r(x)r(y)\rangle_{\mathrm{var}}$
at coinciding points $x=y$.
A standard way to resum  these tadpoles is to use \textit{normal products}.
This procedure consists in replacing 
any monomial $\mathcal{P}[r(x)]$
of the field $r(x)$ at a single point $x$ by the corresponding
normal product (or normal ordered operator) $:\!\mathcal{P}[r(x)]\! :$,
defined 
such that the expectation values of any products of normal ordered operators 
$\langle:\!\mathcal{P}_1\! :\cdots
:\!\mathcal{P}_Q\! :\rangle_{\mathrm{var}}$
at non-coinciding points $x_1\neq\cdots\neq x_Q$ is equal to the sum
over all Feynman diagrams \textit{without tadpoles} which appear in the
evaluation of
$\langle\mathcal{P}_1\cdots
\mathcal{P}_Q\rangle_{\mathrm{var}}$.
All normal ordered operators can be obtained%
\footnote{by differentiating with respect to $\vec k$}
 from the normal ordered 
exponential, satisfying the relation
\begin{equation}
\rme^{\rmi\vec k\vec r(x)}\ =\ \rme^{-{\vec k^{\,2}\over 2}\Cvar}
\,:\!\rme^{\rmi\vec k\vec r(x)}\!: \ ,
\,\label{npexp}
\end{equation}
where $\Cvar$ is the tadpole diagram amplitude
\begin{equation}
\Cvar\ =\ G_{\mathrm{var}}(0)\ =\ 
\int {\rmd^Dq\over (2\pi)^D}\,{1\over q^2+m_{\mathrm{var}}^2}
\ =\ 
\mvar^{D-2}\,\Cc
\label{Ctadpole}
\end{equation}
with
\begin{equation}
\Cc\ =\ (4\pi)^{-D/2}\,\Gamma(1-D/2) \ .
\label{Cc}
\end{equation}
$\Cvar$ coincides with the factor of $A$ in \Eq{equA} of Sect.~4.
Since $\Cvar$ depends on the mass $\mvar$ in 
\Eq{Ctadpole} which is chosen to be the mass appearing in $\hvar$, the
normal ordered products $:\![\cdots]\!:$ depend explicitly on a mass scale
$m$, and should be denoted by $:\![\cdots]\!:_m$. This mass scale
dependence will be omitted in this section, since we shall always choose
$m=\mvar$.

With these notations, we can rewrite the operators in
$\mathcal{H}_V$ and $\hvar$ in terms of normal products.
This gives
\begin{equation}
(\vec r(x))^2\ =\ d\,\Cvar\,\mathbf{1}
\,+\,:\!(\vec r(x))^2\!:
\label{npr2}
\end{equation}
with  the identity operator $\mathbf{1}$.
For $V(\vec r)$ we have
\begin{eqnarray}
V(\vec r)\ &=&\ \int{\rmd^d\vec k\over (2\pi)^d}\,
\tilde V(\vec k)\,\rme^{-{\vec k^{\,2}\over 2}\Cvar}\ 
:\!\rme^{-\rmi\vec k\vec r}\!:
\nonumber\\
&=&\ \sum_{m=0}^\infty\,{(-\rmi)^m\over m!}\,
\int {\rmd^d\vec k\over (2\pi)^d}\,\rme^{-{\vec k^{\,2}\over 2}\Cvar}
\,\tilde V(\vec k)\ :\!\big(\vec k\vec r\big)^m\!:
\ .\label{npV}
\end{eqnarray}
If we make the additional assumption that $V(\vec r)$ 
\textit{is rotational invariant}, then $\tilde V(\vec k)$ depends only on
$|\vec k|=k$ and 
only the even terms $m=2n$ are non zero in \Eq{npV}.
Integrating over $\vec k$ and after some algebra we obtain 
\begin{equation}
V(\vec r)\ =\ \sum_{n=0}^\infty\,
\left(-{1\over 4}\right)^n\,{\Gamma(d/2)\over\Gamma(n+d/2)}\,
{\mathcal{M}_n \over n!}
\,:\!{\big(\vec r^{\,2}\big)}^{n}\!:
\label{seriesV}
\end{equation}
with the moments $\mathcal{M}_n$ given by
\begin{eqnarray}
\mathcal{M}_n\ 
&=&\ \int {\rmd^d\vec k\over (2\pi)^d}\,
\big(\vec k^{\,2}\big)^{n}\,\rme^{-{\vec k^{\,2}\over 2}\Cvar}\,
\tilde V(\vec k)
\nonumber\\
&=&\ 2\,{(4\pi)^{-d/2}\over\Gamma(d/2)}\int_0^\infty \!\rmd k\,k^{d+2n-1}\,
\rme^{-{k^2\over 2}\Cvar}\,\tilde V(k) \ .
\label{Momentn}
\end{eqnarray}
Thus we finally can write  \Eq{equVhat} for $\tilde V(k)$ as
\begin{equation}
\tilde V(k)\ +\ \rme^{-{k^2\over 2}\Cvar}\,
\langle :\!\rme^{\rmi\vec k\vec r(x_0)}\!:\ \rme^{\int_x\Delta}
\rangle_{\mathrm{var}}^{\mathbf{C}}\ =\ 0 \ ,
\label{npeq4V}
\end{equation}
with the ``perturbation'' $\Delta(\vec r)$ written in terms
of normal products as
\begin{eqnarray}
\Delta (\vec r) &=&
 \left[{d\over 2}\,\mvar^2\,\Cvar\,-\,\mathcal{M}_0\right]\,\mathbf{1}
\ +\ \left[{1\over 2}\,\mvar^2\,+\,{1\over 2d}\,\mathcal{M}_1\right]
\,:\!\vec r^{\,2}\!:
\nonumber\\
&\ &-\ \sum_{n=2}^\infty\,\left({-1\over 4}\right)^{n}
\,{\Gamma(d/2)\over\Gamma(n+d/2)}\,{\mathcal{M}_n\over n!}\,
:\!\big(\vec r^{\,2}\big)^n\! : \ .
\label{np4Delta}\end{eqnarray}

Let us first evaluate $\Delta(\vec r)$ when we simply take for $V$ the variational
estimate $V_{\mathrm{var}}$ given by \Eqs{equvarV} and \eq{equA}
\begin{equation}
\tilde V (k) \ \to \ \tilde V_{\mathrm{var}}(k)\ =\ -\,\rme^{-{k^2\over 2}\Cvar} \ .
\label{Vhatvar2}
\end{equation}
We get for the moments
\begin{equation}
\mathcal{M}_{n}\ \to\ 
\mathcal{M}_{n}^{\mathrm{var}} \ = \ 
-\,{\Gamma(n+d/2)\over\Gamma(d/2)}\,\big(4\pi\Cvar\big)^{-d/2}\,\Cvar^{-n}
\label{Mnvar}
\end{equation}
and we can use the self-consistent equation \eq{eqmvar} for $\mvar$
which is in  our notation
\begin{equation}
2\,\mvar^2\,\Cvar\ =\ \big(4\pi\Cvar\big)^{-d/2}
\label{eqmvarCvar}
\end{equation}
to get for $\Delta(\vec r)$
\begin{eqnarray}
\Delta(\vec r) \ \to\ \Delta_{\mathrm{var}}(\vec r)\ &=&\ \mvar^2\,\Cvar\ \left[
\Big({{d\over 2}+2}\Big)\,\mathbf{1}\ +\ 
\sum_{n=2}^{\infty}\, \left({-1\over 4\Cvar}\right)^{n}\,{2\over n!}\,
:\!\big(\vec r^{\,n}\big)^{2}\!: 
\right]\ . \qquad
\label{expDvar}
\end{eqnarray}
The coefficient of the $:\!\vec r^{\,2}\!:$ term is zero, since
$\mathcal{M}_{1}=-d\,\mvar^2$ .

This means that when we take $V_{\mathrm{var}}$ as potential $V$,
$\Delta$ is indeed an interaction term, which contains no
``mass renormalization", but only higher order interaction terms.

\subsection{A convenient rescaling}
To check whether these terms are unimportant in the large $d$ limit,
it is better to rescale all quantities in terms of the variational
mass $\mvar$, that we take as unit of scale.
Thus let us rescale 
\begin{equation}
\begin{array}{rclcrcl}
x\ &\to&\ (\mvar)^{-1}\,x&\qquad&q\ &\to&\ \mvar\,q 
\\
\vec r\ &\to&\ (\mvar)^{D/2-1}\,\vec r&\qquad&\vec k\ &\to&
\ (\mvar)^{1-D/2}\,\vec k
\\
V\ &\to&\ (\mvar)^{D}\,V&
&&&
\end{array}
\label{simplresc} 
\end{equation}
In these units, all previous results obtained in this section are given by
the same equations\footnote{except of course \Eq{eqmvarCvar}.},
provided that we replace
\begin{equation}
\mvar\ \to\ 1\qquad,\qquad\Cvar\ \to\ \Cc
\ ,\label{varto1}
\end{equation}
that we \textit{do not rescale} $\tilde V$ in \Eq{npeq4V}, and that we
rescale the free energy density $\mathcal{E}[V]$ as
\begin{equation}
\mathcal{E}[V]\ \to\ \mvar^{D}\,\mathcal{E}[V] \ .
\label{rescaleE}
\end{equation}
Thus we got rid of the complicated $d$ and $D$ dependence of $\mvar$ and keep
only the simple  factor of $\Cc$ given by \Eq{Cc} in the calculations.

\subsection{The variational solution as $D\to 2$}

In order to estimate ``how close" the variational potential
$V_{\mathrm{var}}$ is  from the exact instanton potential, let us consider
the instanton equation \eq{npeq4V}, where we replace
$V$ by $V_{\mathrm{var}}$ and $\Delta$ by $\Delta_{\mathrm{var}}$.
Then of course \Eq{npeq4V} is not satisfied, since
\begin{equation}
\langle :\!\rme^{\rmi\vec k\vec r(x_0)}\!:\ \rme^{\int_x\Delta_{\mathrm{var}}}
\rangle_{\mathrm{var}}^{\mathbf{C}}\ \neq\ 1 \ .
\label{neq4var}
\end{equation}
This would be true only if all the $n\ge 2$ terms in the expansion
\eq{expDvar} for $\Delta_{\mathrm{var}}$ were zero.

\begin{figure}[t]
\unitlength=1cm
\begin{center}
\noindent
\begin{picture}(7,4)
\put(0,1){\epsfbox{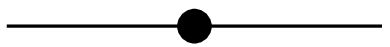}}
\put(5,1){$0$}
\end{picture}
\qquad
\begin{picture}(7,4)
\put(.5,.5){\epsfbox{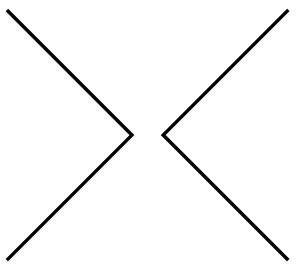}}
\put(5,1){$\displaystyle{1\over 16\,\Cc}$}
\end{picture}
\\
\begin{picture}(7,4)
\put(0,0){\epsfbox{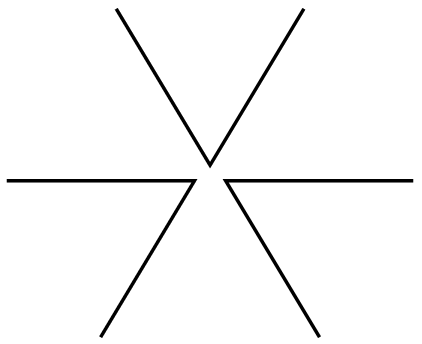}}
\put(4.5,1){$\displaystyle-\,{1\over 192\,\Cc^2}$}
\end{picture}
\qquad
\begin{picture}(7,4)
\put(0,0){\epsfbox{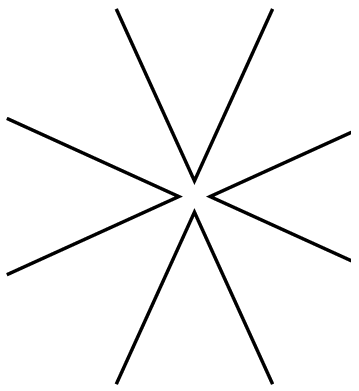}}
\put(4.5,1){$\displaystyle\,{1\over 3072\,\Cc^3}$}
\end{picture}
\end{center}
\caption{The interaction vertices from $\Delta(\vec r)$ in the 
variational approximation.}
\label{f.vertexvar}
\end{figure}

\begin{figure}[t]
\unitlength=1cm
\begin{center}
\begin{picture}(16,5)
\put(0.6,1.5){\includegraphics[scale=1.]{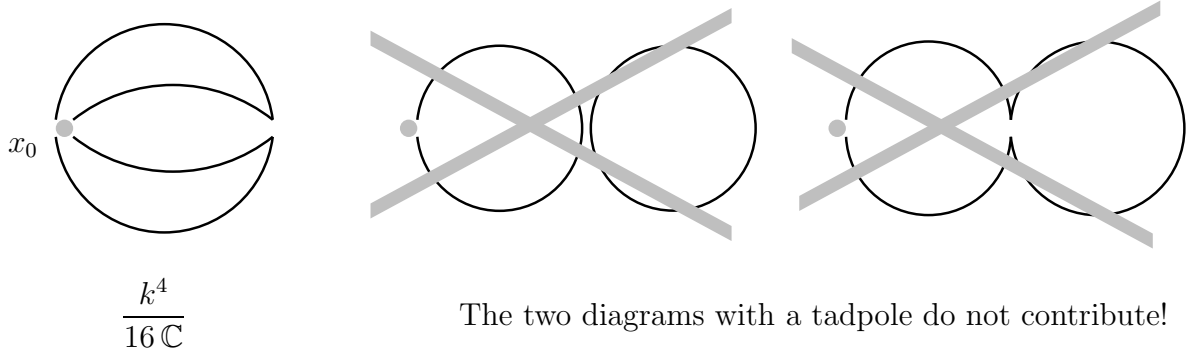}}
\put(1.5,.5){{$\displaystyle{k^4\over 16\,\Cc}$}}
\put(0,2.8){{$\displaystyle x_0$}}
\put(6,.5){The two diagrams with a tadpole do not contribute!}
\end{picture}
\end{center}
\caption{First order graphs in the variational approximation.}
\label{f.1stordergraph}
\end{figure}

\begin{figure}[t]
\unitlength=1cm
\begin{center}
\begin{picture}(16,4.5)
\put(0.6,1.5){\includegraphics[scale=1.]{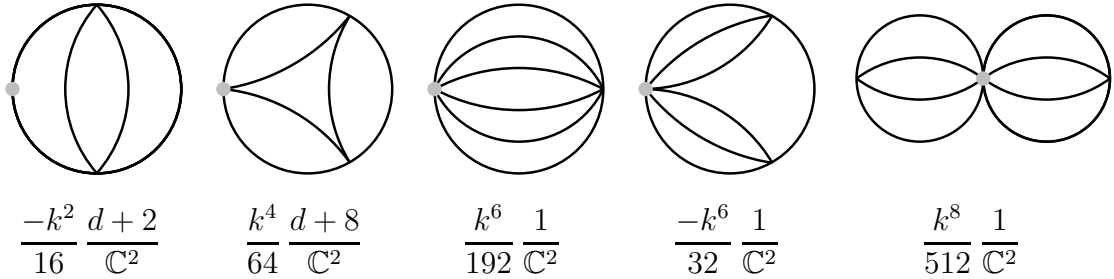}}
\put(.8,.5){{$\displaystyle{-k^2\over 16}\,{d+2\over\Cc^2}$}}
\put(3.8,.5){{$\displaystyle{k^4\over 64}\,{d+8\over\Cc^2}$}}
\put(6.7,.5){{$\displaystyle{k^6\over 192}\,{1\over\Cc^2}$}}
\put(9.5,.5){{$\displaystyle{-k^6\over 32}\,{1\over\Cc^2}$}}
\put(12.8,.5){{$\displaystyle{k^8\over 512}\,{1\over\Cc^2}$}}
\end{picture}
\end{center}
\caption{Second order graphs in the variational approximation.}
\label{f.2ndordergraph}
\end{figure}

We can compute the l.h.s.\ of \Eq{neq4var} in perturbation theory.
The first interaction vertices with their coefficients are depicted on 
\Fig{f.vertexvar}.
Let us note that the 2-point vertex,
corresponding to the $:\!\vec r^{\,2}\!:$ coefficient in \Eq{expDvar}, is zero.
The ``0-point'' vertex, corresponding to the coefficient of
the identity operator $\mathbf{1}$,
 disappears in the connected correlation functions
(since by definition $\langle[\cdots]\,\mathbf{1}\rangle^{\mathbf{C}_{\mathrm{var}}}=0$),
but will be present in the free energy density $\mathcal{E}$.
The propagator is simply $(q^2+1)^{-1}$.
The 0-th order term in the expansion of \Eq{neq4var} is
\begin{equation}
\langle :\!\rme^{\rmi\vec k\vec r(x_0)}\!:\rangle\ \ =\ 1
\label{Oordervar}
\end{equation}
because all tadpoles are subtracted by the normal product prescription.
On \Fig{f.1stordergraph} we have depicted the only diagram which appears
at order $\Cc^{-1}$ with its combinatorial weight resulting from its
symmetry factor, the interaction vertex coefficients and the contractions
of the $d$-dimensional indices of the $\vec r$'s.
We have also depicted the other possible diagrams which do not contribute,
since they contain a tadpole which is subtracted by the normal order
prescription.
The diagrams  at order $\Cc^{-2}$ are depicted on
\Fig{f.2ndordergraph}, together with their weight. 
In general, these diagrams contain internal closed loops, which give a
factor of $d$, and 
open chains which must end at some $\vec k\vec r(x_0)$
in the exponential $:\!\exp\big(\rmi\vec k\vec r(x_0)\big)\!:$, thus giving
a factor of $\vec k^{\,2}$.

We can now look at the limit when the internal dimension of the membrane $D$ goes to $2$, while $\epsilon$ is  fixed.
In this limit, the bulk dimension of space $d$ goes to infinity as
\begin{equation}
d\ =\ {2(2D-\epsilon)\over 2-D}\ \simeq\ {8-2\epsilon\over 2-D}
\label{dsc2-D}
\end{equation}
and the tadpole amplitude $\Cc$ given by \Eq{Cc} diverges like $d$, since
\begin{equation}
\Cc\ =\ (4\pi)^{-D/2}\,\Gamma(1-D/2)\ \simeq\ {1\over 2\pi}\,{1\over 2-D}
\ \simeq\ d\,{1\over 4\pi(4-\epsilon)}\ .
\label{Ccscd} 
\end{equation}
On the other hand, the Feynman amplitude of the first diagram depicted on
\Fig{f.1stordergraph} and of those of \Fig{f.2ndordergraph} are finite when
$D\to 2$, since they do not have any long distance (infra-red) or 
short distance (ultra-violet) divergences for $0\le D\le 2$.
This implies that as $D\to 2$, the contributions of diagrams that we
are considering vanish at least as fast as $2-D\simeq 1/d$.

This result is in fact valid for \textit{all the diagrams} which appear
at higher orders in the evaluation of the l.h.s.\ of \Eq{neq4var}.
All diagrams vanish individually at least as $1/d$, and it is possible to
generate a systematic expansion in powers of $1/d$.
Of course there will be an infinite number of diagrams which contribute
at a given order in $1/d$, that we shall characterize later.
Let us assume that the sum of all diagrams which contribute at a given order
is convergent. (This turns out to be true at least as long as $\epsilon>0$.)
Then this implies that the saddle point equation \eq{npeq4V}
holds at leading order in $1/d$ for the variational solution
\begin{equation}
\tilde V_{\mathrm{var}}(k)\ +\ \rme^{-{k^2\over 2}\Cvar}\,
\langle :\!\rme^{\rmi\vec k\vec r(x_0)}\!:\ \rme^{\int_x\Delta_{\mathrm{var}}}
\rangle_{\mathrm{var}}^{\mathbf{C}}\ =\ 
\rme^{-{k^2\over 2}\Cvar}\ \mathcal{O}\Big({1\over d}\Big)\ ,
\label{eq4Vvar}
\end{equation}
when we take the limit $(2-D)\sim d^{-1}\to 0$, $\epsilon$ and $k$ fixed.

\begin{figure}[t]
\unitlength=1cm
\begin{center}
\begin{picture}(12,4)
\put(1,1.5){\includegraphics{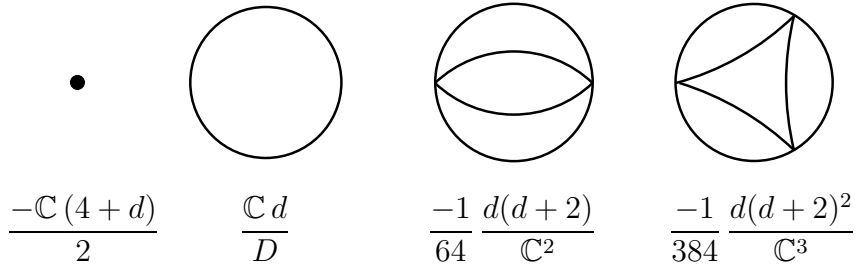}}
\put(0.2,0.5){{$\displaystyle{-\Cc\,(4+d)\over 2}$}}
\put(3.3,0.5){{$\displaystyle{\Cc\,d\over D}$}}
\put(5.8,0.5){{$\displaystyle{-1\over 64}\,{d(d+2)\over\Cc^2}$}}
\put(9.0,0.5){{$\displaystyle{-1\over 384}\,{d(d+2)^2\over\Cc^3}$}}
\end{picture}
\end{center}
\vspace{-5mm}

\caption{Free energy density.}
\label{f.vaccuumdiag}
\end{figure}

A similar statement can be made for the vacuum diagrams which appear in the perturbative expansion of the free energy density.
The first terms of this  expansion are depicted on 
\Fig{f.vaccuumdiag}.
The first two graphs denote the tree and one-loop contributions
given respectively by (minus) the first term in the expansion of $\Delta_{\mathrm{var}}$
\begin{equation}
\mathcal{E}_{\mathrm{var}}^{(0)}\ =\ -\,\Cc\,\Big(2+{d\over 2}\Big)
\label{Evar0l}
\end{equation}
and by the loop integral
\begin{equation}
\mathcal{E}_{\mathrm{var}}^{(1)}\ =\ 
{d\over 2}\,\int{\rmd^Dq\over (2\pi)^D}\ \log\left[1+1/q^2\right]
\ =\ {d\over D}\,\Cc \ .
\label{Evar1l}
\end{equation}
The sum of these two terms gives the variational free energy density
\begin{equation}
\mathcal{E}_{\mathrm{var}}\ =\ \Cc\,\left(-2-{d\over 2}+{d\over D}\right)
\ =\ -\,{\epsilon\over D}\,\Cc
\label{Evarnorm}
\end{equation}
which is of order $\mathcal{O}(d)$, while higher order corrections are at least
of order $\mathcal{O}(1)$ (in our normalization where $\mvar=1$).

What are the consequences of this remarkable fact?
First it will be possible to solve the saddle point equation
\eq{npeq4V} order by order in $1/d$, and to show that the
exact instanton potential $V$ differs from the variational solution
only by  corrections of order $1/d$,
\begin{equation}
\tilde V(k)\ =\ \tilde\Vvar(k)\,\big[1+\mathcal{O}(d^{-1})\big]
\end{equation}
which are finite as long as $\epsilon>0$.
Using this fact we can show that the variational instanton action, which is
in our normalizations
\begin{equation}
\mathcal{S}_{\mathrm{var}}\ =\ 
\mathcal{E}_{\mathrm{var}}\,+\,{1\over 2}\,\int \Vvar^2
\ =\ \left(1-{\epsilon\over D}\right)\,\Cc \ ,
\label{Svarnorm}
\end{equation}
is of order $\mathcal{O}(d)$, and
differs from the exact instanton action by subdominant terms of order
$\mathcal{O}(1)$.
This justifies the use of the variational method as well as
the large order analysis of the $\epsilon$-expansion results made in
the previous Section.

\subsection{Leading $1/d$ correction for the instanton potential}

\begin{figure}
\unitlength=1cm
\begin{center}
\begin{picture}(11.5,4)
\put(1,1.5){\includegraphics{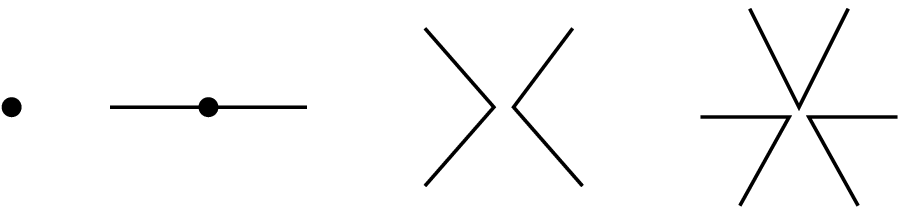}}
\put(0,.5){{$\Cc\left({\displaystyle{d\over 2}}-2\mu_0\right)$}}
\put(3,.5){{$\displaystyle{\delta_1\over 2d}$}}
\put(6,.5){{$\displaystyle{-\mu_2\over 16\Cc}$}}
\put(9,.5){{$\displaystyle{\mu_3\over 192\Cc^2}$}}
\end{picture}
\end{center}
\caption{The vertices from $\Delta(\vec r)$, \Eq{normDelta}.}
\label{f.intver}
\end{figure}

In order to study the $1/d$ corrections, let us rewrite the normal product expansion \eq{seriesV} for the exact instanton potential $V$ as
\begin{equation}
V(\vec r)\ =\ 2\Cc\,\sum_{n=0}^\infty\,\left({-1\over 4\Cc}\right)^{n}\,
{\mu_n \over n!}\,:\!\big(\vec r^{\,2}\big)^{n}\!: \ .
\label{normserV}
\end{equation}
With our rescaling in \Eq{simplresc}, starting from \Eq{Momentn}, the moments $\mu_n$ are given by
\begin{equation}
\mu_n\ =\ {2\,\Cc^{n+d/2}\over\Gamma(n+d/2)}\int_0^\infty\rmd k\,
k^{d+2n-1}\,\rme^{-k^2\Cc/2}\,\tilde V(k) \ .
\label{defmun}
\end{equation}
We assume that $V$ differs from $\Vvar$ by $\mathcal{O}(d^{-1})$, or
equivalently that 
\begin{equation}
\mu_{n}\ =\ -\,1\,+\,{\delta_n\over d}\quad,\quad \delta_n\ =\  \mathcal{O}(1)
\ .
\label{mun}
\end{equation}
The perturbation $\Delta(\vec r)$ given by \Eq{np4Delta} is then
\begin{equation}
\Delta(\vec r)\ =\ \Cc\,\left({d\over 2}-2\mu_0\right)\,+\,
{\delta_1\over 2\,d}\,:\!\vec r^{\,2}\!:\,+\,
(-2\Cc)\,\sum_{n=2}^\infty\,\left({-1\over 4\Cc}\right)^{n}\,
{\mu_n \over n!}\,:\!\big(\vec r^{\,2}\big)^{n}\!:
\label{normDelta} \ .
\end{equation}
This generates the $2n$-point interaction vertices for the
perturbative expansion around the variational Hamiltonian depicted on
\Fig{f.intver}.

The first  vertex with $n=0$ is just the correction to
the free energy.
The additional 2-point vertex is a mass correction and is of order $1/d$.
The $2n$-point vertices ($n\ge 2$) are similar to those of \Fig{f.vertexvar}
and are of order $1/d^{n-1}$.

Before embarking onto the detailed calculations, let us recall what 
we are about to do. Our equation for the {\it exact} instanton potential
$V$ is a self-consistent equation (since both sides contain $V$), and 
was written as a perturbative expansion with respect to the 
variational solution as
\be
- \tilde V(k) \,\rme^{\Cc \,k^2/2 } = \left< 
:\! \rme^{i k r(x_0)}\!: \rme^{\int_x \Delta}
\right>_{\mathrm{var}}^{\mathbf{C}}\ .
\label{recall}
\ee
 $\Delta$, given in \Eq{normDelta}, contains all terms 
of the exact potential $V$ with the exception of those 
 already taken into account in the variational Hamiltonian.

\begin{figure}[b]
\unitlength=1cm
\begin{center}
\begin{picture}(12,2.5)
\put(0,1.3){\includegraphics{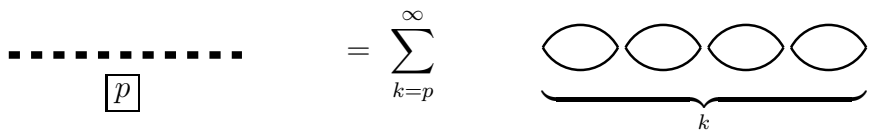}}
\put(4.6,1.5){{$\displaystyle =\ \sum\limits_{k=p}^\infty$}}
\put(7.2,1.1){{$\displaystyle\underbrace{\hspace{4.3cm}}_k$}}
\put(1.4,1.0){\fbox{$\displaystyle p$}}
\end{picture}
\end{center}\vspace*{-5ex}
\caption{The chain of bubbles}
\label{f.chain}
\end{figure}

It is then easy to see that the perturbative expansion generates 
diagrams, which can be organized in a $1/d$ expansion
in terms of the chain of bubbles depicted on \Fig{f.chain}.
Indeed, each $4$-point vertex carries a factor of $\Cc\sim 1/d$ and each bubble
carries a factor of $d$, so that the whole chain is of order $1/d$.
A careful but simple analysis shows that
only four different classes of diagrams contribute to the r.h.s.\ of
\Eq{recall} at order $1/d$; they are depicted on \Fig{1/dgraphs}.
The suffix \fbox{$n$} ($n=0,1$) refers to the minimal number of bubbles
in the chain in order not to have tadpoles.
The situation is thus quite similar to the $1/n$ expansion in models with
a $n$-component field with $O(n)$ symmetry.

\begin{figure}
\begin{center}
\Gb\qquad\Gp{0}\qquad\Gd{1}\\
\vskip3.ex\ \\
\Ge{0}{1}\qquad\Gn{0}{2}\qquad\Gc{0}
\end{center}
\caption{$1/d$ diagrams for the expansion of 
$\langle :\!\rme^{\rmi\vec k\vec r(x_0)}\!:\ 
\rme^{\int_x\Delta_{\mathrm{var}}}\rangle_{\mathrm{var}}^{\mathbf{C}}$}
\label{1/dgraphs}
\end{figure}

The amplitude for a single bubble diagram with external momentum $p$ is
\begin{equation}
\raisebox{-.3truecm}{\includegraphics{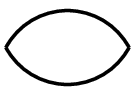}}\ =\ 
\mathbb{B}(p)\ =\ \int{\rmd^Dq\over (2\pi)^D}\,{1\over q^2+1}\,
{1\over (p+q)^2+1}\ =\ {\Gamma(2-D/2)\over (4\pi)^{D/2}}\,\mathbb{J}(p)
\label{bubbleB}
\end{equation}
with
\begin{equation}
\mathbb{J}(p)\ =\ \int_0^1 \rmd x \left[ 1+ x(1-x) p^2 \right] ^{\frac D2 -2}
\label{bubbleJ}
\end{equation}
and can be expressed in terms of hypergeometric functions.
In particular one has
\begin{equation}
\mathbb{B}(0)\ =\ (1-D/2)\,\Cc \ .
\label{Bzero}
\end{equation}
Taking into account the symmetry factor, the amplitude for the 
\fbox{$n$}\,-truncated chain of \Fig{f.chain} is the geometric function
\begin{equation}
\Gj{n}\ =\ 
\mathbb{H}^{(n)}(p)\ =\ \sum_{m\ge n}
\left[(-\mu_2)\,{d\over 4\Cc}\,\mathbb{B}(p)\right]^m
\label{chainH}
\end{equation}
so that in particular the untruncated chain is
\begin{equation}
\Gj{0}\ =\ 
\mathbb{H}^{(0)}(p)\ =\ 
\left[1\,+\,\mu_2\,{2D-\epsilon\over 4}\,\mathbb{J}(p)\right]^{-1} \ .
\label{chainH0}
\end{equation}

\begin{figure}[t]
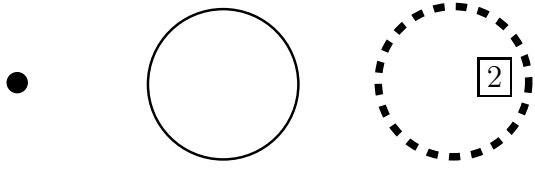

\begin{center}
\Ga\qquad\qquad\Gm\qquad\Gi{2}
\end{center}
\vskip-1ex
\caption{$1/d$ diagrams for the expansion of the free energy density $\mathcal{E}$ at order
$\mathcal{O}(1)$}
\label{1/dvacgraphs}
\end{figure}

Similarly, the free energy density $\mathcal{E}$ can be expanded in $1/d$
in terms of diagrams involving chains of bubbles.
The diagrams which are present at order $\mathcal{O}(1)$ are
depicted on \Fig{1/dvacgraphs}

The last diagram is the \fbox{$2$}\,-truncated closed chain of bubbles,
whose amplitude is (taking into account the symmetry factors)
\begin{equation}
\Gi{n}
\ =\ \mathbb{G}^{(n)}\ =\ \int{\rmd^Dp\over (2\pi)^D}\,
\sum_{m\ge n} {1\over m}\, 
\left[(-\mu_2)\,{d\over 4\Cc}\,\mathbb{B}(p)\right]^m
\label{Gndef} \ .
\end{equation}
With these notations, we now compute the $1/d$ correction for $V$.
With the symmetry factors for the diagrams of \Fig{1/dgraphs},
\Eq{npeq4V} for $\tilde V$ is at that order
\begin{eqnarray}
\tilde V(k)\ &=&\ -\ \rme^{-{k^2\over 2}\Cc}\,
\left[
1
\ +\ 
\Gb\ {- k^2 \delta_1\over 2\,d}
\ +
\right.\nonumber\\
&+&\ \left.\Gp{0}\ {k^2\,\mu_2\,\delta_1\over 8\,\Cc}
\ +\ 
\Gd{1}\ {k^2 \mu_2\over 4\Cc}
\ +
\right.\nonumber\\
&+&\ 
\Gc{0}\ \ \ {- k^4\,\mu_2\over 16\,\Cc}\ +\ 
\Ge{0}{1}\ {-k^2\,d\,\left(\mu_2\right)^2\over 16\,\Cc^2}
\nonumber\\
&+&\ \left. \Gn{0}{2}\ \ \ {k^2\,\mu_3\over 8\,\mu_2\,\Cc}
\ +\ \mathcal{O}\left({1\over d^{2}}\right)\,\right]
\label{1/dhatV} \ .
\end{eqnarray}
Since we are interested in computing $\tilde V$ at order $1/d$ only, and since the diagrams on the r.h.s.\ of \Eq{1/dhatV} are of order $\mathcal{O}(1)$,
we can use the ansatz \eq{mun} and replace $\mu_2$ and $\mu_3$ by $-1$
on the r.h.s.\ of \eq{1/dhatV} (but keep $\delta_1$).
Now we use \Eq{Bzero}, which implies that \\
\begin{equation}
\Gb\ \ =\ (4-\epsilon)\,{\Cc\over d}\ +\ \mathcal{O}\left({1\over d}\right)
\label{Gbzero} \ ,
\end{equation}
and from \Eq{chainH} we have 
\begin{equation}
\Gj{0}\Bigg|_{p=0}\ =\ {4\over\epsilon}\ +\ \mathcal{O}\left({1\over d}\right)
\label{Gjzero}
\end{equation}
We can also rewrite
\begin{equation}
\Gc{0}\ \ =\ \Gg{2}\ \ \left({4\,\Cc\over -d\,\mu_2}\right)^2
\label{Gczero} \ .
\end{equation}
We obtain
\begin{eqnarray}
\tilde V(k)\ &=&\ -\ \rme^{-k^2\Cc/2}\,\left[\,
1\ +\ (-k^2)\,\left(\delta_1\,{2\,(4-\epsilon)\,\Cc\over \epsilon\,d^2}\ +\ 
\Gd{1}\ {1\over\epsilon\,\Cc}\ +\right.\right.
\nonumber\\
&+&\ \left.\left.
\Gg{2}\ {\epsilon-4\over 2\,d\,\epsilon}
\right)
\ +\ 
(k^4)\ \Gg{2}\ {\Cc\over d^2}\  
\ +\ \mathcal{O}\left({1\over d^2}\right)
\right] \ .
\label{V1/d-2}
\end{eqnarray}
Now we have to compute $\delta_1$.
It is fixed by \Eq{defmun} which relates the $\mu_n$'s to $\tilde V$.
Indeed we have from \Eqs{defmun} and \eq{mun}
\begin{equation}
\delta_1\ =\ d\,+\,d\,
{2\,\Cc^{1+d/2}\over\Gamma(1+d/2)}\int_0^\infty\rmd k\, k^{d+1}\,
\rme^{-k^2\Cc/2}\,\tilde V(k)
\label{eqdel1-1}
\end{equation}
and using \Eq{V1/d-2}, performing the $k$-integral and keeping only the leading
terms as $d\to\infty$, we obtain the equation
\begin{equation}
\delta_1\ =\ {(4-\epsilon)\over \epsilon}\,\delta_1\ +\ 
\Gd{1}\ {d^2\over 2\,\Cc^2\,\epsilon}\ +\ \Gg{2}\ {(-d)\over\Cc\,\epsilon}
\ +\ \mathcal{O}\left({1\over d}\right) \ .
\label{eqdel1-2}
\end{equation}
Finally, we can use the simple relation
(derived in appendix~\ref{a.1})
\begin{equation}
D\ \Gi{2}\ =\ 
(4-D)\,\Gh{2}\ +\ {\mu_2\,d\over\Cc}\Gk{1}
\label{3graphsid}
\end{equation}
to express $\delta_1$ in terms of the simple closed chain of bubbles
$\mathbb{G}^{(2)}$ defined by \Eq{Gndef}
\begin{equation}
\delta_1\ =\ 
{d\over 2\,\Cc\,(2-\epsilon)}\ 
\Gi{2}
\ +\ \mathcal{O}\left({1\over d}\right)
\label{delta1fin} \ .
\end{equation}
Inserting this result into \Eq{V1/d-2}, we obtain the potential $\tilde V$,
and we can calculate the moments $\mu_n$ from \Eq{defmun}.
Since for $n\ll d$, the moments are independent of $n$, we find that
\begin{equation}
\mu_n\ =\ -1\,+\,{\delta_1\over d}
\ +\ \mathcal{O}\left({1\over d^2}\right)
\qquad\forall\,n\ge 0 \ .
\end{equation}
Therefore the ansatz \eq{mun} is consistent and our calculation at order
$1/d$ makes sense, provided that the closed chain of bubbles $\mathbb{G}^{(2)}$,
which is a function of $D$, $\epsilon$ and $\mu_2$, has a finite limit
as $D\to 2$ (and $\mu_2= -1$).
This is the case as long as $\epsilon>0$.
Indeed, the chain $\mathbb{G}^{(2)}$ is given by \Eq{Gndef} and using
\Eqs{bubbleB} and \eq{bubbleJ}, it is given in the limit $D\to 2$ by
\begin{eqnarray}
{G}(\epsilon)\ &=&\ \left.\Gi{2}\right|_{D=2}
\nonumber\\
&=&\ \int {\rmd^2q\over (2\pi)^2}\,\left[
-\,\log\left[1-\left(1-{\epsilon\over 4}\right)J(q)\right]\,-\,
\left(1-{\epsilon\over 4}\right)J(q)\right]
\label{G2D2}
\end{eqnarray}
where
\begin{eqnarray}
J(q)\ &=&\ \left.\mathbb{J}(q)\right|_{D=2}\ =\ 
\int_0^1\rmd x\,\left[1+x(1-x)q^2\right]^{-1}
\nonumber\\
&=&\ {2\over q\,\sqrt{q^2+4}}\,
\log\left[{\sqrt{q^2+4}+q\over\sqrt{q^2+4}-q}\right]
\label{JD2} \ .
\end{eqnarray}
Indeed, we have for $J(q)$
\begin{equation}
0<J(q)<1\ \forall\,q\ne 0\ ;\ 
J(q)\,\simeq\,1\,-\,{q^2\over 6}\ \mathrm{as}\ q\to 0\ ;\ 
J(q)\,\simeq\,{2\over q^2}\,\log(q^2)\ \mathrm{as}\ q\to\infty
\label{jbehavior} \ ,
\end{equation}
hence the integral \eq{G2D2} defining the function $G(\epsilon)$ is 
convergent for any $\epsilon\ge 0$.

Similarly, the chain with a marked point is given in the limit $D\to 2$ by
\begin{eqnarray}
\overline{G} ({\epsilon})\ &=&\ \left.\Gh{2}\right|_{D=2}
\nonumber\\
&=&\ \int {\rmd^2 q\over (2\pi)^2}\,\left[
{1\over 1-\left(1-{\epsilon\over 4}\right)\,J(q)}\,-\,1\,-\,
\left(1-{\epsilon\over 4}\right)\,J(q)
\right]
\nonumber\\
&=&\ (4-\epsilon)\,{\rmd\over\rmd\epsilon}G(\epsilon)
\label{barG2D2} \ .
\end{eqnarray}
It is finite for $\epsilon>0$, but has a logarithmic divergence at
$\epsilon=0$, since from \eq{jbehavior} the integrant in \eq{barG2D2} behaves
as $q^{-2}$ as $q\to 0$ when $\epsilon=0$.
A simple calculation shows that $\bar G$ has a logarithmic singularity of the form
\begin{equation}
\bar G(\epsilon)\ \simeq\ {3\over 2\pi}\,\log\left({1\over\epsilon}\right)
\qquad\mathrm{as}\qquad \epsilon\to 0
\label{singbarG}
\end{equation}

With these notations, the final result for the instanton potential
is at order $1/d$
\begin{eqnarray}
\tilde V(k)&=&\rme^{-k^2\Cc/2}\left[
 -1+{1\over d}\left( k^2\left({1\over 2-\epsilon}G(\epsilon)+
{4-\epsilon\over 2}G'(\epsilon)\right)
-k^4{1\over 4\pi}G'(\epsilon)
\right)+
\mathcal{O}\left({1\over d^2}\right)
\right] \ .
\nonumber\\
\label{finalV1/d} 
\end{eqnarray}
We note that the corrections to the variational potential are finite
as long as $0<\epsilon<2$, but are singular at $\epsilon=0$ and $\epsilon=2$.
The singularity at $\epsilon=2$ is not surprising.
We have seen from \Eq{mvarfinal} that the variational mass $\mvar$ itself
is singular at
$\epsilon=D$, which corresponds in the limit $D\to 2$ to $\epsilon=2$.
Therefore it is expected that the corrections to the variational result
will also be singular.
The singularity at $\epsilon=0$ is new, since $\mvar$ is regular at
$\epsilon=0$.
We shall come back to this issue later.

\subsection{$1/d$ correction to the instanton action}

It is now easy to compute the corrections to the instanton action $\mathcal{S}$,
defined by \Eq{actioninst}.
The free energy density $\mathcal{E}$ has the following graphical expansion
\begin{eqnarray}
\mathcal{E}(V) &=& -\,\Cc\,\left({d\over 2}-2\,\mu_0\right)\,\Ga\ 
+\ {d\over 2}\,\Gm\ -\ {1\over 2}\,\Gi{2}
\,+\,\mathcal{O}\left({1\over d}\right)\quad 
\label{freeen1/d}
\end{eqnarray}
where the amplitude for the simple loop is given by \Eq{Evar1l}
\begin{equation}
\Gm\ =\ \int {\rmd^D q\over (2\pi)^D}\,\log\left[1+1/q^2\right]
\ =\ {2\over D}\,\Cc
\label{loop2} \ .
\end{equation}
The second term in \Eq{actioninst} is
\begin{eqnarray}
{1\over 2}\int \rmd ^d\vec r\,V(\vec r)^2\ &=&\ 
{1\over 2}\,\int{\rmd^d\vec k\over(2\pi)^d}\,|\tilde V(\vec k)|^2 \ .
\end{eqnarray}
Within our rescaling \eq{simplresc} this term is 
\begin{equation}
{1\over 2}\int V^2\ \to\ \Cc\,{2\,\Cc^{d/2}\over\Gamma(d/2)}\,\int_0^\infty\rmd k\,
 k^{d-1}\,|\tilde V(k)|^2
\label{v2resc}
\end{equation}
and from \Eqs{defmun} and \eq{mun} it is simply obtained from the moment
$\mu_0$ by
\begin{equation}
{1\over 2}\int V^2\ =\ \Cc\,\Big[-1\,-\,2\,\mu_0\,+\mathcal{O}(d^{-2})
\Big] \ .
\label{v21/d}
\end{equation}
Thus the instanton action, which is the sum of \Eqs{freeen1/d} and
\eq{v21/d} is finally given at order $1/d$ by
\begin{eqnarray}
\mathcal{S}(V) &=& \Cc\ \left({d\over 2}+4\right)\,\Ga\ -\ {d\over 2}\,
\Gm\,-\,{1\over 2}\,\Gi{2}\,+\,\mathcal{O}\left({1\over d}\right)
\\
&=&\ 
\Cc\,\left(1-{\epsilon\over D}\right)\,-\,{1\over 2}\,\mathbb{G}^{(2)}(D,\epsilon)
\,+\,\mathcal{O}\left({1\over d}\right)
\label{instact1/d} \ .
\end{eqnarray}
This result holds within the rescaling \eq{simplresc}.
The first term on the r.h.s.\ of \Eq{instact1/d} is  the
variational result, and is of order $\mathcal{O}(d)$.
The second term is of order  $\mathcal{O}(1)$.
Thus the first correction to the variational result is
given by the closed loop diagram of \Eq{G2D2}.

According to \Eq{rescaleE}, to recover 
the normalization used in the previous sections, we have simply to
multiply \Eq{instact1/d} by $\mvar^D$, where $\mvar$ is the variational mass
as given by \Eq{mvarfinal}.

\subsection{Discussion of the result}
\label{s.6.7}
\noindent
(1) The corrections of the variational instanton potential do in fact not 
contribute to the correction of the instanton action. Indeed, they are 
entirely contained in the moment $\mu_0$ which disappears in
\Eq{instact1/d}.
This is not surprising, since the instanton potential $V$ is defined by
a variational principle with respect to $V$.\\
\ \\
\noindent
(2) The correction to the instanton action is finite as $\epsilon\to 0$,
while from \Eq{finalV1/d} the correction to the potential $V$ diverges
as $\log(\epsilon)$.
The divergences occur in fact because the chain of bubbles of \Fig{f.chain},
$\mathbb{H}^{(0)}(q)$, has a singularity at small momentum $q$ as $D\to2$ and
$\epsilon\to 0$.
According to \Eqs{chainH0} and \eq{jbehavior}, it behaves as
\begin{equation}
\left.\Gj{0}\,\right|_{q\to0}\ \simeq\ {1\over q^2/6+\epsilon/4+{\cal O}(2-D)} 
\label{H0div} \ .
\end{equation}  
\noindent
In other word, this chain of bubbles behaves as a massless propagator at
$\epsilon=0$ and $D=2$.
We have no physical interpretation of these IR singularities, and why they
cancel in the instanton action.
However, the fact that as $D\to 2$ the instanton potential is singular as
$\epsilon\to 0$ is similar to the situation at $D=1$ discussed in
section 5.1. 
Here also, we know from the exact solution that the instanton potential
is singular when $\epsilon \to 0$ ($d\to 4$), while the instanton action is
regular, and in fact \Eq{r0dto4} corresponds also to a logarithmic
behavior as $\epsilon\to 0$.
Therefore, it is reasonable to conjecture that at
next orders in $1/d$, the IR singularities which occur
as $\epsilon\to 0$ in the instanton potential still disappear in the instanton action.
This would imply that the coefficient $\mathcal{C}$ has a regular behavior
when $\epsilon\to 0$ and that the limits $d\to\infty$ and $\epsilon\to 0$
can be exchanged, providing further consistency to our arguments for
the large order behavior of the $\epsilon$-expansions for the SAM model.
\\
\ \\
\noindent
(3) The $1/d$ correction to the instanton action $\mathcal{S}_{\mathrm{inst}}$
is negative, as expected, since it should improve the variational estimate
for $\mathcal{S}_{\mathrm{inst}}$, which is an upper bound. \\
\begin{figure}[t]
\unitlength=1.cm
\begin{center}
\begin{picture}(11,8)
\put(0,1){\includegraphics{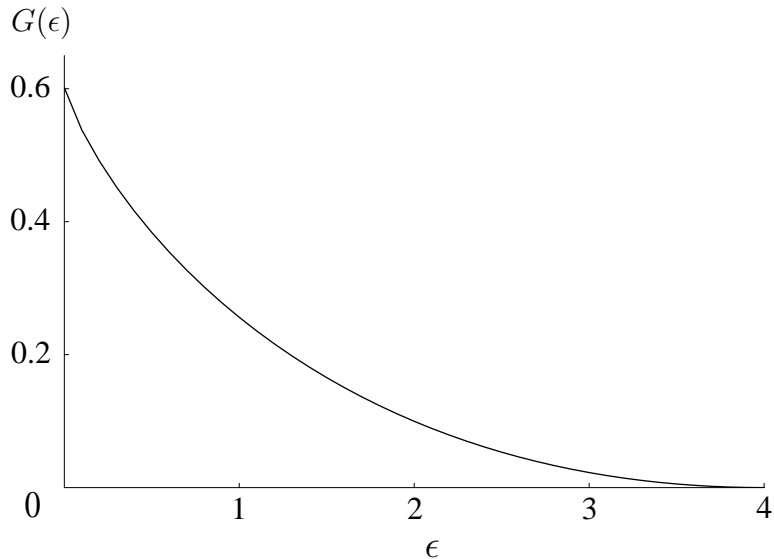}}
\put(5.5,.5){\large $\epsilon$}
\put(0,7.5){$G(\epsilon)$}
\end{picture}
\end{center}
\vskip-5.5ex
\caption{The function $G(\epsilon)$ defined in \Eq{G2D2} from the chain
of bubbles in $D=2$.}
\label{f.Geps}
\end{figure}
\\ 
\noindent
(4) It is interesting to see if the $1/d$ correction can be used to improve 
the variational estimates for quantities of physical significance, such as
the large order constant $\mathcal{C}$.
Let us start from the result \eq{instact1/d} for the instanton action $\mathcal{S}$ at first order in $1/d$, and insert it into \Eq{largenz}
for $\mathcal{C}$. 
We obtain
\begin{equation}
\mathcal{C}^{-1}\ =\
\mathcal{C}_{\mathrm{var}}^{-1}\,\left[1\,-\,{\mathbb{G}^{(2)}(D,\epsilon)
\over 2\,\Cc(D)}\,+\,\mathcal{O}(d^{-2})\right]
\ .
\label{SLGW1/d}
\end{equation}
Now we only keep  the leading $1/d$ corrections on the r.h.s.\ of \Eq{SLGW1/d}
in the limit $d\to\infty$, for  fixed $\epsilon$ by using
\begin{equation}
\mathbb{G}^{(2)}(D,\epsilon)\ =\ G(\epsilon)\,+\,\mathcal{O}(d^{-1})
\ ,\qquad
\Cc(D)\ =\ {d\over 4\,\pi}\,{1\over4-\epsilon}\,+\,\mathcal{O}(1)
\ ,
\label{G2C1/d}
\end{equation}
with $G(\epsilon)$ given by \eq{G2D2}.
We thus obtain
\begin{equation}
\mathcal{C}^{-1}\ =\
\mathcal{C}_{\mathrm{var}}^{-1}\,\left[1\,-\,
\,2\pi(4-\epsilon)G(\epsilon)\,{1\over d}\,+\,\mathcal{O}(d^{-2})\right]
\ .
\label{SLGW1/d2}
\end{equation}
We have plotted on \Fig{f.Geps} the function $G(\epsilon)$ for $0<\epsilon<4$,
as obtained by a straightforward numerical integration.
$G$ is maximal for $\epsilon=0$, where we have $G(0)={0.6014\ldots}$.

Let us estimate the first $1/d$ correction for the case of polymers in
$d=4$ dimensions.
We simply have to set $d=4$ and $\epsilon=0$ in \eq{SLGW1/d2}.
We find that the $1/d$ correction is
\begin{equation}
2\pi(4-\epsilon)G(\epsilon)\,{1\over d}\ \to\ 2\,\pi\,G(0)\ =\ {3.78\ldots}
\ ,
\label{D=1corr}
\end{equation}
which is much larger than $1$ !
Thus the $1/d$ correction is very large for $d=4$ and one should take into
account the subleading corrections and resum them.
We recall that for $d=4$ and $\epsilon=0$ we have 
$\mathcal{C}_{\mathrm{exact}}^{-1}/\mathcal{C}_{\mathrm{var}}^{-1}=2/3=0.666\ldots$.  In practice, we expect that for $\epsilon=0$ the $1/d$ corrections
will be smaller or equal to the variational $\mathcal{O}(d^0)$ result
for $d>16$ !
Therefore we cannot use naively the $1/d$ corrections that we have
calculated to improve the variational estimates.
In the next subsection we propose an improved resummation procedure,
which takes into account some of the higher order corrections which
are contained in the bubble chain diagrams, and which gives much better 
results.

\subsection{An attempt to go beyond the first $1/d$ correction}\label{s6.8}
In the last section we have shown that a straightforward $1/d$-expansion
cannot be applied to small dimension, as e.g.\ polymers in $d=4$. 
In this section, we propose a different approximation scheme. It consists
in summing exactly the chain of bubbles $\mathbb{G}^{(2)}$, and keeping 
the full $D,d$-dependence when extrapolating to low dimensions, instead of
expanding this quantity about $d=\infty$.
First, in \Eq{Gndef}, only the leading $d$-dependence had been
kept. Taking into account the complete $d$-dependence, we obtain
\begin{equation}
\Gi{n}
\ =\ \mathbb{G}^{(n)}\ =\ \frac{d}{d+2}\int{\rmd^Dp\over (2\pi)^D}\,
\sum_{m\ge n} {1\over m}\, 
\left[(-\mu_2)\,{d+2\over 4\Cc}\,\mathbb{B}(p)\right]^m
\label{Gndefcomplete} \ .
\end{equation}
Eliminating as before $\mathbb{B}(p)$ in favor of  $\mathbb{J}(p)$,
$d$ in favor of $\E$ and replacing $\mu_2$ by its leading 
contribution $\mu_2=-1$ (justified later),
we obtain for the correction to the 
variational result
\bea \label{delta S improved}
\frac{ S-S_{\mathrm{var}}}{S_{\mathrm{var}}} &=& 
\frac{D(2D-\E)}{(2+D-\E)(D-\E) }
\frac{\sin{\frac{\pi D}2}}{\pi} \nn\\
&& \qquad \times \int_0^\infty \rmd p \, 
p^{d-1}\left[\ln\left(1-\frac{2+D-\E}4 \mathbb{J}(p)\right)
+\frac{2+D-\E}4 \mathbb{J}(p)\right] \ ,\qquad
\eea
In the remainder, we shall focus on the case
 $D=1$, for which we can most easily test \Eq{delta S improved}.
In $D=1$, $\mathbb{J}(p)$ is exactly given by 
\be
\mathbb{J}(p)=\frac{4}{4+p^2} \ .
\ee
\Eq{delta S improved} is then  integrated (using the residue calculus) with 
the result
\be \label{delta S for small d}
\left.\frac{ S-S_{\mathrm{var}}}{S_{\mathrm{var}}}\right|_{D=1}
=\frac{2 d}{(d+2)(d-2)}\left( \sqrt{3-\frac d2}+\frac d8 -\frac74\right)\ .
\ee
\begin{figure}[t]
\unitlength=1.cm
\begin{center}
\begin{picture}(11,8)
\put(0,-1.3){\includegraphics{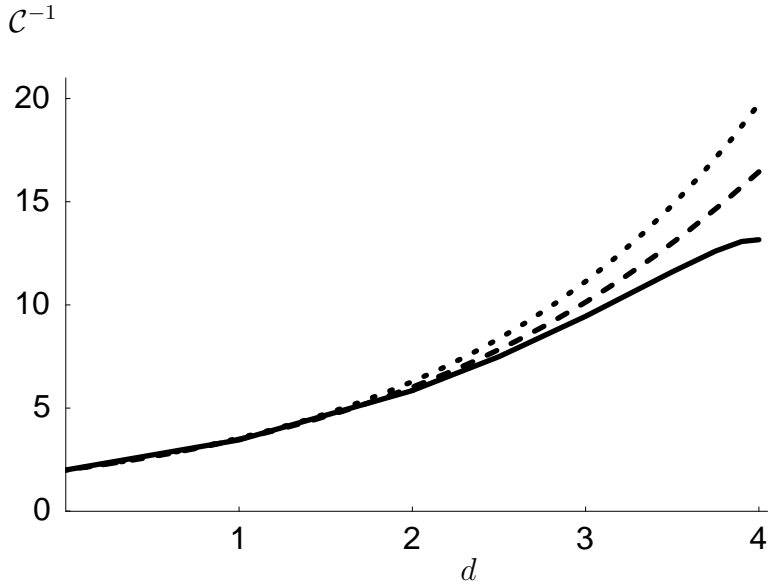}}
\put(6,.3){{$d$}}
\put(0,7.6){{$\displaystyle\mathcal{C}^{-1}$}}
\end{picture}
\end{center}
\vskip-4ex
\caption{The inverse of the large order constant $1/\mathcal{C}$ for
the Edwards model ($D=1$) as a function of the bulk dimension $d$.
The dotted curve is the $\mathcal{O}(1)$ variational estimate, the dashed curve the  estimate from \Eq{C for small d}, the continuous curve the exact result.
}
\label{f.D1plot}
\end{figure}%
The large order estimate is finally obtained as 
\begin{equation} \label{C for small d}
\mathcal{C}^{-1}(D=1)\ \simeq\ 2\,\pi^{d/2}\ 
\left[ 1
\,+\,
\frac d{d+2} \left(\sqrt{3-\frac d2}+\frac d8 -\frac74\right)
+\,\ldots\right] \ ,
\end{equation}
which is  plotted on \Fig{f.D1plot}.
We see that 
this corrects 50\% of the deviation of the variational result 
from the exact result in $d=4$, and is even better in lower dimensions. 
Note that this is not the straightforward $1/d$-correction, obtained 
in section \ref{s.6.7}.
In the following, we want to justify that the result given in 
\Eqs{delta S for small d} and \eq{C for small d} is
meaningful for small dimension, by analyzing the case $D=1$ and 
$d=4$, i.e.\ $\E=0$, where calculations are most easily done. 

The first observation is that $\mathbb{H}^{(0)}(p)$ is nicely 
convergent for all values of $p$. 
In contrast to $D=2$, no divergence at $p=0$ appears. 

The next observation is that all terms in the chain of bubbles 
$\mathbb{G}^{(2)}$ are positive
and the sum rapidly converging. (If we denote by 100\% the 
difference between variational estimate and exact result, 
the contributions of the first terms are 28\%, 11\%, 4\%,\ldots,
respectively, which add up to the 50\% given above.)

We also observe that most of the missing contributions to 
$S(V)$ are positive:
Let us call a vertex even, if it possesses an 
 even number of pairs of lines, and odd otherwise. Any even 
vertex then contributes a factor of 1, whereas any odd vertex contributes
a factor of $(-1)$ (see \Eqs{normserV} and \eq{mun}). Since all integrals
are positive, in order 
to build up a negative contribution, an odd number of odd vertices
has to be taken. One sees by inspection, that this is more difficult 
than for a positive diagram (containing even vertices and an even number of 
odd vertices). The first diagram of this kind is
\be
\includegraphics{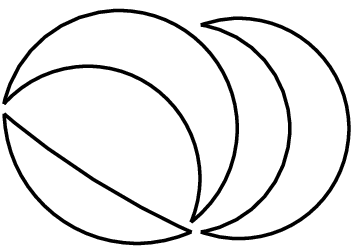}
\ee

Another hint comes from a second class of diagrams, which can be 
summed, namely all ``watermelon" diagrams with 2 vertices only:
\be
\includegraphics{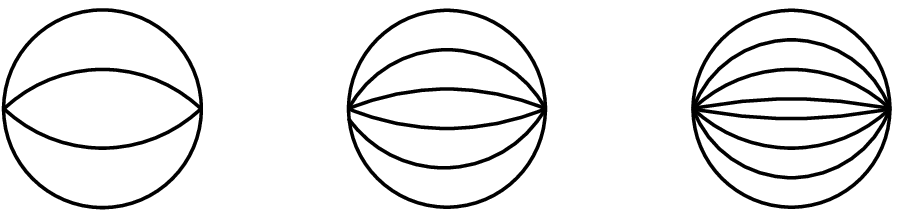}
\ee
The first diagram, equivalent to  28\%, is already contained in 
$\mathbb{G}^{(2)}$, the following contribute 6\%, 1.5\%, \ldots 
respectively. They can be resummed and are then given by the integral
\be
\int_0^\infty\rmd x\, \left(1-\frac{\rme^{-2x}}4 \right)^{-d/2}-1-\frac d8\rme^{-2x}\ ,
\ee
which contributes about 8.2\% in $d=4$, when neglecting the first 
diagram already taken into account in $\mathbb{G}^{(2)}$.

A more systematic approach to understand why higher order
vertices are subdominant, is to look for an additional small 
parameter. We have seen that ${\mathbb{C}}\sim d$ for $d\to \infty$.
Comparing the ratio ${d}/{\mathbb{C}}$ at $d=\infty$ and 
$d=4$, we obtain (always for $\E=0$)
\be
\frac{\displaystyle\left.{d}/{\mathbb{C}}\right|_{d=4}}
{\displaystyle\left.{d}/{\mathbb{C}}\right|_{d=\infty}}
=\frac1{2\pi} \ .
\ee
This suggests (as exemplified by  the preceding calculation) 
that higher order vertices, which come with additional factors of 
$1/\mathbb{C}$, indeed contribute an additional small factor.

The last point to verify, is that also the $1/d$-correction 
to the potential is small. In the notation of \Eqs{normserV} and \eq{mun},
the parameter $\delta_1/d$, given in \Eq{delta1fin}, reads  
\be
\left.\frac{\delta_1}d \right|_{D=1,\E=0}\ =\ \frac1{4\mathbb{C}}\  
\left.\Gi{2}\right|_{D=1,\E=0}
\hskip-1.5em +\ {\cal O}\left(\frac1{d^2}\right)\ 
=\frac1{12} +{\cal O}\left(\frac1{d^2}\right)
\ee
It is indeed small. 

Another important limit to verify is $d\to0$. We have argued in 
section \ref{s.5.1} that in that limit the variational result
becomes exact. Our correction in \Eq{delta S improved} consistently 
vanishes. Diagramatically this can be seen from the 
fact that {\em any} diagram which contributes to the free energy
has at least one closed loop, and therefore comes with a factor
of $d$. This is also the leading contribution for small $d$: due to
 the ``islands and bridges theorem'', which states
that if from any island an even number of bridges starts, it is
always possible to construct a path which uses any bridge exactly
once, the minimal number of closed loops, and thus factors of 
$d$, for any given diagram is one. It is therefore impossible
to simply reorganize the perturbative expansion in a $d$-expansion. 
However, the large-$d$ expansion should respect this exact 
property. This is not assured by the naive $1/d$-expansion, 
but by our modified result discussed here. 

All these arguments suggest that \Eq{C for small d} is a 
sensible correction to  the variational result, and 
should be improved by taking into account higher 
order vertices, although this expansion is not a systematic expansion 
in a small parameter. 


\section{Conclusion}
\label{s.7}

Let us first summarize the results of this paper.
We have developed a general formulation to estimate the large orders of perturbation theory for the Edwards model of self-avoiding membranes and polymers.
We have shown that these large orders are controlled by a semi-classical
effective potential $V(\vec r)$, solution of a non-local extremization
problem.
This effective potential is the analog 
to the
Lipatov instanton in Landau-Ginsburg-Wilson models.
In the case of polymers (membranes with internal dimension $D=1$), this
SAM-instanton corresponds precisely to the Lipatov instanton for the
$n=0$ components LGW model.
The large order behavior for the SAM model is derived in the general case
($D\ne 1$).
The equations for the SAM-instanton are solved within a Gaussian variational
approximation.
The result is for $D=1$ compared  to the exact results from Lipatov's method,
and is found to be  qualitatively correct for polymers in $d=4$ dimensions.
Finally it is shown that the variational result is the first term of a
systematic expansion in $1/d$, where $d$ is the dimension of space in which
the membrane fluctuates.

All these results are new, and represent a considerable advance in the
understanding of self-avoiding membranes beyond the first orders of
perturbation theory.
A number of interesting issues still has to be addressed:\\
\noindent
- We have seen that a systematic expansion in $1/d$ can be constructed,
of which the variational approximation is only the leading term.
However, adding the first correction in $1/d$ to the
 instanton action does not give reliable results for small $d$. 
This is a numerical problem which comes from the fact that in the interesting cases, the corrections in $1/d$ are already quite large.
We have proposed a modified procedure to resum the first corrections
which gives better results when applied to the case of polymers
($D=1$, $d=4$), but more systematic resummation methods are needed in order to
improve in a reliable way the variational results.\\
\noindent
- We have not calculated the contributions from the fluctuations around the
instanton, which should give the value of the global constant 
$\mathcal{A'}$ in \Eq{asympz}.
This calculation is  technically more difficult than the one for the
instanton action itself.\\
- We have only discussed shortly and at a qualitative level the
consequences of our results for the $\epsilon$-expansion of
the scaling exponents.
It would be interesting to obtain more precise results.
In particular, the discussion relies on the limit $\epsilon\to 0$ for the
large order estimates. 
We have shown that this limit exists in the variational approximation, but
that the $1/d$ corrections then suffer from IR divergences.
These divergences cancel for the instanton action at first order in $1/d$,
and we conjectured that this feature persists at higher orders.
Further studies are needed to clarify this important problem.\\
\noindent
- Finally, let us mention that our approach can be applied to other classes
of interactions between membranes.\\

\section*{Acknowledgments:}
F.~D. thanks J.~des Cloizeaux  
for a useful discussion.
K.~W.\ grateful acknowledges useful discussions with H.W.~Diehl. It is
equivalently a 
pleasure for him to thank
L.~Sch\"afer for numerous enlightening discussions  at an 
early stage of this work. We thank E.~Guitter
 for his interest and proof reading.

\appendix
\section{The $\Phi^4$ instanton in the limit $d\to 4$}
\label{a.2}
The instanton equation \eq{eqins1D} for $\Psi_0$ is, assuming rotational
invariance i.e.\ $\Psi_0(\vec r)=\Psi_0(r)$ with $r=|\vec r|$,
\begin{equation}
\Psi_0''\,+\,{d-1\over r}\,\Psi_0'\,+\,2\,E_0\,\Psi_0\,+\,2\,\Psi_0^3\ =\ 0
\ .
\label{insteqapp}
\end{equation}
We look for the solution which is regular at the origin, which has no zero
for finite $r$ and which vanishes at infinity, i.e.
\begin{equation}
0<\Psi_0(r)<\infty\quad \mathrm{for}\quad 0\le r<\infty\quad,\quad
\Psi_0(r)\to 0\quad \mathrm{when}\quad r\to\infty
\ .
\end{equation}
$E_0\le 0$ is fixed by the normalization \eq{normpsi0}
\begin{equation}
\|\Psi_0\|^2\ =\ 
\int \rmd^d\vec r\,\Psi_0(\vec r)^2\ =\ {2\,\pi^{d/2}\over\Gamma(d/2)}
\int_0^\infty\rmd r\,r^{d-1}\,\Psi_0(r)^2\ =\ 1
\ .
\label{NormCond}
\end{equation}
From \Eq{ELGW} we know that $E_0$ can be extracted from the action of
the instanton for the LGW action \eq{SLGW}, which is finite when $d\to 4$,
such that $E_0$ behaves as
\begin{equation}
E_0\ \propto\,d-4
\ .
\label{E0d24}
\end{equation}
For $d=4$ and $E_0=0$, the general solution of \Eq{insteqapp} is
not normalizable and reads
\begin{equation}
\Psi_0(r)\ =\ {2\,r_0\over r^2+r_0^2}
\ .
\end{equation}
For $d\ne 4$, $E_0<0$ and $r\to\infty$, $\Psi_0$ is a solution of the
linearized equation $\Psi''+(d-1)/r\Psi'+2E_0\Psi=0$, given by a Bessel
function
\begin{equation}
\Psi_0(r)\ \simeq\ r^{1-d/2}\,K_{d/2-1}\left(|2E_0|^{1/2}\,r\right)
\ .
\end{equation}
Assuming that
\begin{equation}
r_0^2\,|E_0|\ \ll\ 1 \qquad\mathrm{when}\quad d\to 4\ ,
\label{condr0E0}
\end{equation}
it has the asymptotics
\be
\Psi_0(r)
 \propto  \left\{
\begin{array}{lll}
 r^{2-d} &\quad \mathrm{for}\quad& r^2\,|2E_0|\,\ll\,1 \\
 r^{{1-d\over 2}}\,\rme^{-|2E_0|^{1/2}\,r} &\quad \mathrm{for}\quad &r^2\,|2E_0|\,\gg\,1
\end{array} \right.
\ \ \ .
\label{BesselAs}
\ee
For $r_0\ll r\ll|2E_0|^{-1/2}$, $\Psi_0$ decays algebraically as
$1/r^2$.
It is exponentially decreasing for $r\gg(-2E_0)^{-1/2}$.
We may thus approximate the integral in \Eq{NormCond} by
\begin{equation}
\|\Psi_0\|^2\ \approx\ 
2\pi^2\int_{r_0}^{{a\over\sqrt{4-d}}}\rmd r\,r^3\ \left({r_0\over r^2}\right)^2
\approx 2\pi^2\ r_0^2\,\log\left[{a\over r_0\sqrt{4-d}}\right]
 \ ,
\end{equation}
where $a$ is the proportionality factor in \Eq{E0d24}.
From the constraint $\|\Psi_0\|=1$ we deduce that the size of the instanton
$r_0$ becomes small as $d\to 4$ and scales as
\begin{equation}
r_0^2\ \sim\ {1\over |\log\left(4-d\right)|}
\ .
\end{equation}
This is consistent with the assumption \eq{condr0E0}.

\section{A simple relation}
\label{a.1}
We derive the relation
\begin{equation}
D\ \Gi{2}\ =\ 
(4-D)\,\Gh{2}\ +\ {\mu_2\,d\over\Cc}\Gk{1}
\ .
\label{relappA}
\end{equation}
For that purpose, we write the amplitude for the last diagram as
an integral over $\vec p$, the $D$-dimensional momentum which flows through the
chain of bubbles
\begin{equation}
\Gk{1}\ =\ 
\int {\rmd^D\vec p\over (2\pi)^D}\,\mathbb{D}(p)\,\mathbb{H}^{(1)}(p)
\ =\ 
{2\,(4\pi)^{-D/2}\over\Gamma(D/2)}\, \int_0^\infty \rmd p\,p^{D-1}\,
\mathbb{D}(p)\,\mathbb{H}^{(1)}(p)
\ .
\label{GkappA}
\end{equation}
The definition \eq{chainH} for the chain of bubbles was
\begin{equation}
\mathbb{H}^{(1)}(p)\ =\ 
\Gj{1}\ =\ 
\left[1+\mu_2{d\over 4\,\Cc}\mathbb{B}(p)\right]^{-1}\,-\,1
\ ,
\label{H1appB}
\end{equation}
and $\mathbb{D}(p)$ is the amplitude for the bubble with a single mass
insertion
\begin{equation}
\mathbb{D}(p)\ =\ \vec p\,\to\,
\raisebox{-.3truecm}{\includegraphics{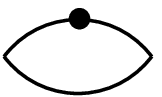}}
\ =\ \int{\rmd^D\vec q\over (2\pi)^D}\,{1\over(\vec q^2+1)^2}\,
{1\over (\vec p+\vec q)^2+1}
\ .
\label{DdefappB}
\end{equation}
We are careful in distinguishing the vector $\vec p$ from its modulus
$p=|\vec p|$.
Similarly the amplitudes for the first two diagrams are
\begin{equation}
\Gi{2}\ =\ {2\,(4\pi)^{-D/2}\over\Gamma(D/2)}\, \int_0^\infty \rmd p\,p^{D-1}\,
\left[-\,
\log\left[1+\mu_2{d\over 4\,\Cc}\mathbb{B}(p)\right]\,+\,
\mu_2{d\over 4\,\Cc}\mathbb{B}(p)
\right]
\ ,
\label{GiappA}
\end{equation}
and
\begin{equation}
\Gh{2}\ =\ {2\,(4\pi)^{-D/2}\over\Gamma(D/2)}\, \int_0^\infty \rmd p\,p^{D-1}\,
\left[
\left(1+\mu_2{d\over 4\,\Cc}\mathbb{B}(p)\right)^{-1}\,-1\,+\,
\mu_2{d\over 4\,\Cc}\mathbb{B}(p)
\right]
\ .
\label{GhappA}
\end{equation}
We can obtain $\mathbb{D}(p)$ from $\mathbb{B}(p)$.
Let us introduce a mass $m$ and consider
\begin{equation}
\mathbb{D}(p,m)
\ =\ \int{\rmd^D\vec q\over (2\pi)^D}\,{1\over(\vec q^2+m^2)^2}\,
{1\over (\vec p+\vec q)^2+m^2}
\label{Dmass}
\end{equation}
and similarly let us introduce a mass in the bubble $\mathbb{B}$ of
\Eq{bubbleB} and define
\begin{equation}
\mathbb{B}(p,m)
\ =\ \int{\rmd^D\vec q\over (2\pi)^D}\,{1\over(\vec q^2+m^2)}\,
{1\over (\vec p+\vec q)^2+m^2}
\ .
\label{Bmass}
\end{equation}
It is easy to see that one has
\begin{equation}
m\,{\rmd\over\rmd m}\,\mathbb{B}(p,m)\ =\ -\,4\,m^2\,\mathbb{D}(p,m) \ ,
\end{equation}
and since by homogeneity
\begin{equation}
\mathbb{B}(p,m)\ =\ m^{D-4}\,\mathbb{B}(p/m) \ ,
\end{equation}
we deduce that (setting at the end $m=1$)
\begin{equation}
(D-4)\,\mathbb{B}(p)\,-\,p\,{\rmd\over\rmd p}\,\mathbb{B}(p)\ =\ 
-\,4\,\mathbb{D}(p)
\ .
\label{D2B}
\end{equation}
We can use \Eq{D2B} to express $\mathbb{D}(p)$ in terms of $\mathbb{B}(p)$ and
its derivative in \Eq{GkappA} and integrate by part in order to
eliminate the derivative of $\mathbb{B}$.
We obtain
\begin{eqnarray}
\Gk{1}\ =\ 
{2\,(4\pi)^{-D/2}\over\Gamma(D/2)}&\,&
\int_0^\infty\!\rmd p\,p^{D-1}\,
\left(1-{D\over 4}\right)\,
\mathbb{B}(p)\,
\left[\left(1+{\mu_2\,d\over 4\,\Cc}\mathbb{B}(p)\right)^{-1}-1\right]
\nonumber\\
&&\,-\,{D\over 4}\,
\left[{4\,\Cc\over \mu_2\,d}\,
\log\left(1+{\mu_2\,d\over 4\,\Cc}\mathbb{B}(p)\right)\,-\,\mathbb{B}(p)
\right]
\ .
\label{Gkpartint}
\end{eqnarray}
Using \Eqs{GiappA} and \eq{GhappA} we obtain the identity \eq{relappA}.


\end{document}